\documentclass[twocolumn,showpacs,preprintnumbers,amsmath,amssymb]{revtex4}

\usepackage{graphicx}
\usepackage{dcolumn}
\usepackage{bm}
\def\upa{\uparrow}
\def\dna{\downarrow}
\def\E{\rm E}
\def\J{\rm J}
\def\bJ{\beta{\rm J}}
\def\S{\rm S}
\def\be{\begin{equation}}
\def\ee{\end{equation}}
\def\bd{\begin{displaymath}}
\def\ed{\end{displaymath}}
\def\-{\phantom{-}}

\begin{document}
%
%

\title{Neutron Scattering and Magnetic Observables for S = 1/2
Spin Clusters and Molecular Magnets}

\author{J.T.Haraldsen}
\affiliation{Department of Physics and Astronomy, University of
Tennessee, Knoxville, TN 37996}

\author{T.Barnes}
\affiliation{Department of Physics and Astronomy, University of
Tennessee, Knoxville, TN 37996 \\
Physics Division, Oak Ridge National Laboratory, Oak
Ridge, TN 37831}

\author{J.L.Musfeldt}
\affiliation{Department of Chemistry, University of Tennessee,
Knoxville, TN 37996}

\begin{abstract}

In this paper we report results for magnetic observables
of finite spin clusters composed of S~=~1/2
ions. We consider clusters of two, three and four spins
in distinct spatial arrangements, with
isotropic Heisenberg interactions of various strengths
between ion pairs.
In addition to the complete set of
energy eigenvalues and eigenvectors,
specific heat and
magnetic susceptibility,
we also quote results for the single crystal and powder average
inelastic neutron scattering structure factors.
Examples of the application of these results to
experimental systems are also discussed.

\end{abstract}


\maketitle

\clearpage

\section{Introduction}

Recent years have seen a rapid increase in the
interest in finite quantum spin systems, also known as
molecular magnets or nanomagnets
\cite{Dag96,Bar99,DiV99,Nie00,Fur00,Bou01,Cor02,Men99,Men00,Cif01}.
Molecular magnets typically consist of clusters
of interacting spins that are magnetically isolated
from the other clusters in the molecular solid by nonmagnetic ligands.
Formally, molecular
magnets are materials in which the ground state has nonzero total
spin. Here we generalize this definition to include
all systems of largely isolated clusters of interacting quantum spins.
These materials are interesting both as simple model
systems for the study of quantum magnetism
and because they have possible applications as nanoscale
computer memory elements
\cite{DiV99,Nie00}.
Many realizations of finite spin clusters
with various ionic spins, ground state spins and geometries
have been reported
in the literature;
some recent examples with S=1/2 ions are given in Table~\ref{materials}.

Theoretical results for the properties of finite S=1/2
quantum spin systems have appeared in several recent references, primarily
in the context of experimental studies of specific materials.
Dimer results are reported in several studies of the
S=1/2 spin dimer
VO(HPO$_4$)$\cdot$0.5H$_2$O;
see for example
Johnson {\it et al.} \cite{Joh84},
Tennant {\it et al.} \cite{Ten97} and
Koo {\it et al.} \cite{Koo04}.
Theoretical properties of S=1/2 spin trimers have similarly been given
in studies of candidate trimer materials; see for example
Refs.\cite{Lub02,Qiu02,Cag03a,Cag03b,Kor04}.

Rather few general
theoretical results have been reported for S=1/2 spin tetramers, since
the results are more complicated and
there are many more independent geometries and sets of superexchanges.
Specific cases of tetramers are considered by
Procissi~{\it et al.} \cite{Pro04} (S=1/2 square tetramer),
Gros~{\it et al.} \cite{Gro03}
and Jensen~{\it et al.} \cite{Jen03} (an unsymmetric S=1/2
tetrahedral model of Cu$_2$Te$_2$O$_5$(Br$_{1-x}$Cl$_x$)),
Kortz~{\it et al.} \cite{Kor04} (unsymmetric tetramer model of
K$_7$Na[Cu$_4$K$_2$(H$_2$O)$_6$($\alpha$-AsW$_9$O$_{33}$)$_2$]$\cdot$5.5H$_2$O),
and
Ciftja \cite{Cif01} (symmetric trimer with apical spin).
More general reviews of quantum spin systems have been
published by
Kahn \cite{Kah93} (thermodynamics) and
Whangbo {\it et al.} \cite{Wha03} (local origins of magnetism,
thermodynamics properties, and materials).
Studies of the dynamics of Heisenberg spin clusters using
a quasiclassical formalism have been reported
in a series of papers by Ameduri, Efremov and Klemm
\cite{Ame02,Efr02,Kle02}. Waldmann \cite{Wal03} has carried out calcuations
of the inelastic neutron structure factor for cyclic Heisenberg spin
clusters which are quite similar to the results presented here.

This increased level of interest in molecular magnets
motivates more detailed theoretical investigation of
the properties of finite quantum spin systems.
For simple theoretical models such as the Heisenberg model,
clusters that consist of only a few interacting
magnetic ions can be treated analytically, and
closed-form results can be obtained for many physical
observables. One especially interesting quantity is the
inelastic neutron scattering structure factor, which is
required for the interpretation of inelastic neutron
scattering experiments.
Inelastic neutron scattering is very well suited to the
investigation of magnetic interactions at interatomic scales,
since the measured structure factor is sensitive to the local
geometry and interactions of the magnetic ions.
As this work is intended in part to facilitate future neutron scattering
studies, the evaluation of this structure factor
is one of our principal concerns.

\begin{table*} [ht]
\caption{Some examples of small S = 1/2 quantum spin systems.}
\vskip 0.5cm
\begin{ruledtabular}
\begin{tabular}{llcl}
Material & Spin System & Ground State S$_{tot}$ & Refs. \\
\colrule
\hline
&   \\
VO(HPO$_4$)$\cdot$0.5H$_2$O
& dimer
& 0 
&
\cite{Joh84,Ten97,Koo04}
\\
Cu$_3$(O$_2$C$_{16}$H$_{23}$)$_6\cdot 1.2$C$_6$H$_{12}$
& symmetric trimer
& 1/2
& \cite{Cag03a,Cag03b}\\
Na$_9$[Cu$_3$Na$_3$(H$_2$O)$_9$($\alpha$-AsW$_9$O$_{33}$)$_2$]$\cdot$26H$_2$O
& symmetric trimer
& 1/2
& \cite{Kor04} \\
$[$Cu$_3$(cpse)$_3$(H$_2$O)$_3]\cdot 8.5$H$_2$O
& symmetric trimer
& 1/2
& \cite{Lop02} \\
(CN$_3$H$_6$)$_4$Na$_2$[H$_4$V$_6$O$_8$(PO$_4$)$_4$((OCH$_2$)$_3$CCH$_2$OH)$_2$]$\cdot
14$H$_2$O
& isosceles trimer
& 1/2
& \cite{Lub02} \\
Na$_6$[H$_4$V$_6$O$_8$(PO$_4$)$_4$((OCH$_2$)$_3$CCH$_2$OH)$_2$]$\cdot
18$H$_2$O
& general trimer
& 1/2
& \cite{Lub02} \\
K$_6$[V$_{15}$As$_6$O$_{42}$(H$_2$O)]$\cdot$8H$_2$O
& symmetric trimer + capping hexamers
& 1/2
& 
\cite{Mue88,Bar92,Gat91,Cha02,Cha04}\\
NaCuAsO$_4$
& linear tetramer
& 0
& \cite{Ulu03,Nag03} \\
(NHEt$_3$)[V$_{12}$As$_8$O$_{40}$(H$_2$O)]$\cdot $H$_2$O.
& rectangular tetramer + capping tetramers
& 0
& \cite{Bas02}\\
K$_{7}$Na[Cu$_4$K$_2$(H$_2$O)$_6$($\alpha$-AsW$_9$O$_{33}$)$_2$]$\cdot$5.5H$_2$O
& distorted tetramer
& 1
& \cite{Kor04}
\\
\label{materials}
\end{tabular}
\end{ruledtabular}
\end{table*}

In this paper we specialize to magnets that are
clusters of S=1/2 ions with isotropic Heisenberg interactions,
and give analytic results for the properties of
dimer, trimer and tetramer clusters with various geometries.
After the introduction, in Sec.II we define the Heisenberg model
and the observables we evaluate in this work.
These include the
standard thermodynamic quantities
for magnetic materials (partition function, specific heat and
magnetic susceptibility), as well as the
inelastic neutron scattering structure factors.
In Section III we evaluate these quantities for specific spin clusters,
which are the spin dimer, symmetric, isosceles and general spin trimers,
and three cases of spin tetramers
(tetrahedral, rectangular, and alternating linear).
We also tabulate all the energy eigenvalues and
eigenvectors for each spin system.
Section III ends with an application,
which is a numerical study of powder inelastic
neutron scattering amplitudes in NaCuAsO$_4$ \cite{Ulu03}; the results appear
to support the identification of this material
with the alternating linear tetramer model and agree well with the
data of Nagler {\it et al.} \cite{Nag03}.
Finally, Section IV discusses several materials
which may be candidates for future experimental studies,
as well as some interesting directions for future theoretical research.

For reference purposes our principal results for the spin systems 
considered here are given in a series of tables at the end of the paper.
These results are 
the spectrum of energy eigenvalues and 
eigenvectors (Table~\ref{wavefunctions1}),
allowed inelastic neutron scattering transitions between these states
(Table~\ref{INSE}), and the specific heats (Table~\ref{C_table})
and susceptibilities (Table~\ref{chi_table}). 

Most previous theoretical studies of molecular magnets in the literature have
specialized to individual materials and their associated model
Hamiltonians. Our results are intended to be sufficiently
general so that they should be useful for the
interpretation of data on many candidate molecular magnets.

\section{The Model and Observables}

\subsection{The Heisenberg Magnet}

The nearest-neighbor Heisenberg magnet, which we shall assume as our
standard model for molecular magnets, is defined by the
Hamiltonian

\be
{\cal H} = \sum_{<ij>}  {\rm J}_{ij}\;  \vec{\rm S}_{i}\cdot
\vec{\rm S}_{j}
\label{magH}
\ee
where the superexchange constants $\{ {\rm J}_{ij}\} $ are positive for
antiferromagnetic interactions and negative for ferromagnetic ones,
and $\vec {\rm S}_i$ is the quantum spin operator for a spin-1/2 ion
at site $i$.

Since this is a rotationally invariant Hamiltonian in spin space,
the total spin S$_{tot}$ is a good quantum number. For the specific cases of
dimer, trimer and tetramer clusters of S=1/2 ions that
we consider here, the energy eigenstates have the total spin decompositions
given below.

\be
1/2 \otimes 1/2 = 1 \oplus 0 ,
\label{spins2}
\ee

\be
1/2 \otimes 1/2 \otimes 1/2 = 3/2 \oplus 1/2 ^2 ,
\label{spins3}
\ee

\be
1/2 \otimes 1/2 \otimes 1/2 \otimes 1/2
= 2 \oplus 1^3 \oplus 0^2 .
\label{CGseries}
\ee
Each S$_{tot}$ multiplet contains
2S$_{tot}+1$ magnetic states, which are degenerate given an isotropic
magnetic Hamiltonian such as the Heisenberg form of Eq.(\ref{magH}).

\subsection{Expressions for Observables}

The energy eigenstates and eigenvalues may be found by
diagonalizing the magnetic Hamiltonian on a convenient basis.
(In practice we will employ
the usual set of $\hat z$-polarized magnetic basis states.)
Several physically interesting quantities may be computed directly from
the energy eigenvalues; in this work these are the 
partition function, specific
heat and magnetic susceptibility, which are given by
\be
Z = \sum_{i=1}^N\, e^{- \beta \E_i}
 = \sum_{\E_i}\, (2{\rm S}_{tot}+1)\, e^{- \beta \E_i} \ ,
\label{partition}
\ee
\be
C =  k_B \beta^{\, 2}\, \frac{d^2 \! \ln (Z)}{d \beta^2} \ ,
\label{C}
\ee
and
\bd
\chi = \frac{\beta}{Z}
\sum_{i=1}^N  \, (M_z^2)_i\, e^{-\beta \E_i}
\ed
\be
=  \frac{1}{3} (g \mu_B)^2 \frac{\beta}{Z}
\sum_{\E_i} \, (2{\rm S}_{tot}+1) \, ({\rm S}_{tot}+1)\,
{\rm S}_{tot}\, e^{- \beta \E_i} \ .
\label{chi}
\ee
In these central formulas
the sum
$i=1\dots N$
is over all $N$ independent energy eigenstates (including magnetic substates),
the sum $\sum_{\E_i}$ is over energy levels only,
$M_z = m g \mu_B$ where
$m = \S_{tot}^z/\hbar $ is the integral or half-integral
magnetic quantum number, and $g$ is the electron $g$-factor.

In addition to these bulk quantities, we also give results for
inelastic neutron scattering intensities. In ``spin-only" magnetic neutron
scattering at zero temperature, the
differential cross section for the inelastic scattering of an incident neutron
from a magnetic system in an initial state $|\Psi_i\rangle$,
with momentum transfer $\hbar\vec q$ and energy transfer
$\hbar\omega$,
is proportional to the neutron scattering structure factor tensor
\bd
S_{ba}(\vec q, \omega) = \hskip 3cm
\ed
\be
\int_{-\infty}^{\infty}\! \frac{dt}{2\pi} \
\sum_{\vec x{_i}, \vec x{_j}}
e^{i\vec q \cdot (\vec x_i - \vec x_j )  +i\omega t}
\langle \Psi_i |
\S_b^{\dagger}(\vec x_j, t) \S_a(\vec x_i, 0)
| \Psi_i \rangle \ .
\label{Sab_def0}
\ee
The site sums
in Eq.(\ref{Sab_def0}) run over all
magnetic ions in one unit cell, and $a,b$ are the spatial indices
of the spin operators.

For transitions between discrete energy levels,
the time integral gives a trivial delta function
$\delta(\E_f - \E_i - \hbar \omega)$
in the energy transfer, so it is useful to specialize to an
``exclusive structure factor" for the excitation of states
within a specific
magnetic multiplet (generically $|\Psi_f (\lambda_f)\rangle $)
from the given initial state
$|\Psi_i \rangle $,

\be
S_{ba}^{(fi)}(\vec q\, ) =
\sum_{\lambda_f}\
\langle \Psi_i |
V_b^{\dagger}
| \Psi_f (\lambda_f)\rangle \ \langle \Psi_f (\lambda_f)|
V_a
| \Psi_i \rangle  \ ,
\label{Sab_def}
\ee
where the vector $V_a(\vec q\,) $ is a sum of spin operators
over all magnetic ions in a unit cell,
\be
V_a = \sum_{{\vec x}_i} {\S}_a(\vec x_i)\;
e^{i\vec q \cdot \vec x_i } \ .
\label{Va_defn}
\ee

This exclusive structure factor is related to
the exclusive differential inelastic neutron scattering cross section by
\be
\frac{\ d\sigma^{(fi)}}{d\Omega} =
(\gamma r_0)^2 \frac{k'}{k} (\delta_{ab} - \hat q_a \hat q_b)
S^{(fi)}_{ba}(\vec q\,) |{\rm F}(\vec q\,)|^2
\label{cross_sec}
\ee
where $\gamma \approx -1.913$ is the neutron gyromagnetic ratio,
$r_o = \hbar \alpha / m_e c$ is the classical electron radius,
$k$ and $k'$ are the magnitudes of the
initial and final neutron wavevectors, and
${\rm F}(\vec q\,)$ is the ionic form factor.
(This relation is abstracted from Eq.(7.61) of Ref.\cite{Squ78},
specialized to an exclusive process.)

For a rotationally invariant magnetic interaction
and an S$_{tot} = 0$ initial state (as is often encountered
in T=0 inelastic scattering from an antiferromagnet), only
S$_{tot}~=~1$ final states are excited, and
$S_{ba}^{(fi)}(\vec q\, ) \propto \delta_{ab}$.
In this case we may
define a scalar neutron scattering structure factor
$S(\vec q\, )$ by
\be
S_{ba}^{(fi)}(\vec q\, ) =  \delta_{ab}\, S(\vec q\, ) \ .
\label{Str1}
\ee

The result for $S_{ba}^{(fi)}(\vec q\, )$
is more complicated for neutron scattering
from a magnetic (S$_{tot} > 0$) initial state.
If we assume an isotropic magnetic Hamiltonian and
a spherical basis for the spin operators S$_a$,
the tensor $S_{ba}^{(fi)}(\vec q\, )$
is diagonal but is not $\propto \delta_{ab}$; it instead has
entries that are proportional to a universal function of $\vec q$
times a product of Clebsch-Gordon coefficients, since
\be
\langle\Psi_f(\lambda_f)|
\, V_a \,
|\Psi_i(\lambda_i)\rangle
=
\langle {\S}_f \lambda_f | 1 a \ {\S}_i \lambda_i \rangle
\ {\cal V}^{(fi)}(\vec q\, ) \ ,
\label{Vred_defn}
\ee
where ${\cal V}^{(fi)}(\vec q\, )$ is the reduced matrix element
for the transition $|\Psi_i\rangle \to |\Psi_f\rangle$.
Here we simplify the presentation by quoting the unpolarized
result $\langle S_{ba}^{(fi)}(\vec q\, )\rangle$,
obtained by summing over final and averaging over initial
polarizations. This unpolarized
$\langle S_{ba}^{(fi)}(\vec q\, )\rangle$
{\it is} $\propto \delta_{ab}$, so it suffices to give the
function $S(\vec q\, )$;
\bd
\langle S_{ba}^{(fi)}(\vec q\, )\rangle  =  \delta_{ab}\, S(\vec q\, )
=
\ed
\be
\frac{1}{2\S_i+1}\sum_{\lambda_i,\lambda_f}\
\langle \Psi_i (\lambda_i) |
V_b^{\dagger}
| \Psi_f (\lambda_f)\rangle\
\langle \Psi_f (\lambda_f)|
V_a
| \Psi_i(\lambda_i) \rangle \ .
\label{Strunpoldefn}
\ee
If desired, the general results for polarized scattering can be recovered by
reintroducing the appropriate Clebsch-Gordon coefficients
of Eq.(\ref{Vred_defn}) in Eq.(\ref{Sab_def}).

The results given above apply to neutron scattering from
single crystals. To interpret neutron experiments on powder samples, 
we require an orientation average of the unpolarized 
single-crystal neutron scattering structure factor.
We define this powder average by
\be
{\bar S}(q) = \int \frac{d\Omega_{\hat q}}{4\pi}\, S(\vec q\, ) \ .
\label{Strpowavg}
\ee

\section{Results for Specific Cases}

\subsection{Spin Dimer}

\begin{figure}[ht]
\includegraphics[height=3in,width=2.5in]{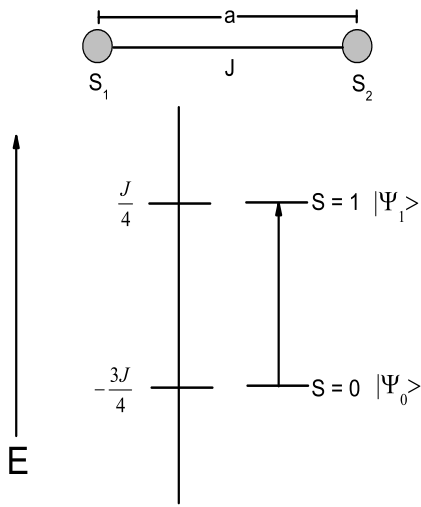}
\caption{The geometry and energy levels of a Heisenberg spin dimer.}
\label{dimerdia}
\end{figure}

The ``minimal" spin cluster model is the S = 1/2 spin dimer
(Fig.\ref{dimerdia}), which consists of a single pair of S = 1/2 spins
interacting through the Heisenberg Hamiltonian,
\be
{\cal H} = {\rm J}\, \vec{\rm S}_{1}\cdot\vec{\rm S}_{2} \ .
\label{dimerH}
\ee
Since this is an isotropic magnetic
Hamiltonian, the total spin is a good quantum number, and
from the Clebsch-Gordon series
$1/2 \otimes 1/2 = 1 \oplus 0$
we expect the spectrum
to consist of an
S$_{tot} = 1$ triplet
and an
S$_{tot} = 0$ singlet.
In a  $\hat z$-diagonal basis
\be
\left[
\begin{array}{l}
|\!\upa\upa\,\rangle
\\
|\!\upa\dna\,\rangle
\\
|\!\dna\upa\,\rangle
\\
|\!\dna\dna\,\rangle
\\
\end{array}
\right]
\ee
the Hamiltonian matrix is
\be
{\cal H} =
{\rm J}
\left[
\begin{array}{cccc}
1/4  &        &       &     \\
     &  -1/4  &  \phantom{-}1/2   &     \\
     &  \phantom{-}1/2  & -1/4  &     \\
     &        &       & 1/4
\end{array}
\right].
\label{dimerHmatrix}
\ee
Diagonalizing this Hamiltonian matrix gives the
energy eigenvalues and eigenvectors,
\be
\begin{array}{l}
{\rm E}_1 = \phantom{-}\frac{1}{4}\,{\rm J}
\\
\\
{\rm E}_0 = -\frac{3}{4}\,{\rm J} \ ,
\end{array}
\label{dimerE}
\ee
\be
\left[
\begin{array}{l}
|\Psi_1(+1)\rangle =
|\!\upa\upa\,\rangle
\\
\\
|\Psi_1(\phantom{+}0)\rangle =
\frac{1}{\sqrt{2}}(|\!\upa\dna\,\rangle + |\!\dna\upa\,\rangle)
\\
\\
|\Psi_1(-1)\rangle =
|\!\dna\dna\,\rangle
\\
\end{array}
\right]
\label{dimervec1}
\ee

\be
|\Psi_0 \rangle =
\frac{1}{\sqrt{2}}(|\!\upa\dna\,\rangle - |\!\dna\upa\,\rangle)
\ .
\quad
\label{dimervec0}
\ee

\begin{figure}[ht]
\includegraphics[width=3.4in]{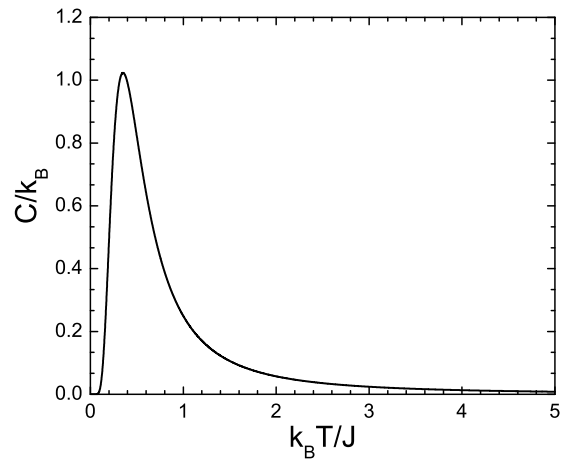}
\caption{The magnetic contribution to the specific heat
of a spin dimer, Eq.(\ref{C_dimer_eq}) (dimensionless units).}
\label{C_dimer}
\end{figure}

\begin{figure}[ht]
\includegraphics[width=3.4in]{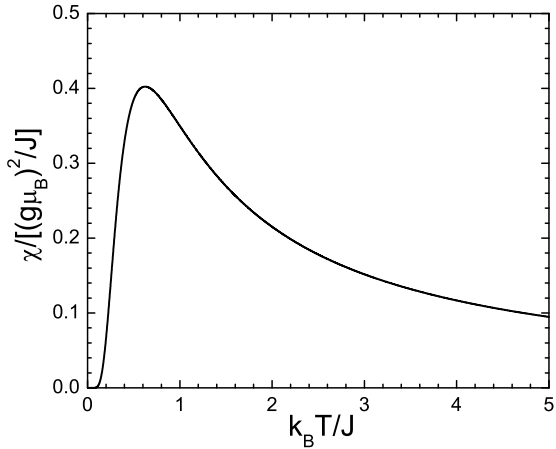}
\caption{The magnetic susceptibility of a spin dimer,
Eq.(\ref{chi_dimer_eq})
(dimensionless units).}
\label{chi_dimer}
\end{figure}

The specific heat and magnetic
susceptibility for the dimer are
especially simple,
since there is only a single
excited level. The results are
(in a dimensionless form)
\be
Z =
e^{\frac{3}{4}\bJ}
+
3 e^{-\frac{1}{4}\bJ}
\ee
\be
\frac{C}{k_B} =
\; 3(\bJ)^2
\frac
{e^{-\bJ}}
{\big(1+3 e^{-\bJ}\big)^2} \ ,
\label{C_dimer_eq}
\ee
and
\be
\frac{\chi}{(g\mu_B)^2/{\J}} =  \,
\; 2 \bJ \,
\frac
{e^{-\bJ}}
{(1+3 e^{-\bJ})}
\ .
\label{chi_dimer_eq}
\ee
These results
are summarized in Tables
\ref{C_table}
and
\ref{chi_table},
as are the corresponding results we find
for the other spin systems considered in this paper.
Plots of the dimensionless specific heat and susceptibility of the
spin dimer are shown in Figs.\ref{C_dimer},\ref{chi_dimer}.

One may confirm that this specific heat formula gives the correct
entropy for a dimer of S=1/2 ions,
\be
S = \int_0^{\infty}\!\! C \; \frac{d\beta}{\beta} = k_B\, 2\ln(2) \ .
\label{Entropy_dimer}
\ee
The corresponding result for a general spin system is
\be
S = k_B\,
\ln({\cal N}/{\cal N}_0) \ ,
\label{Entropy_general}
\ee
where ${\cal N}$ is the dimensionality of the full Hilbert space
and ${\cal N}_0$ is the degeneracy of the ground state; for the
S=1/2 dimer, ${\cal N} = 2^2$ and ${\cal N}_0=1$.

\begin{figure}[ht]
\includegraphics[width=3.4in]{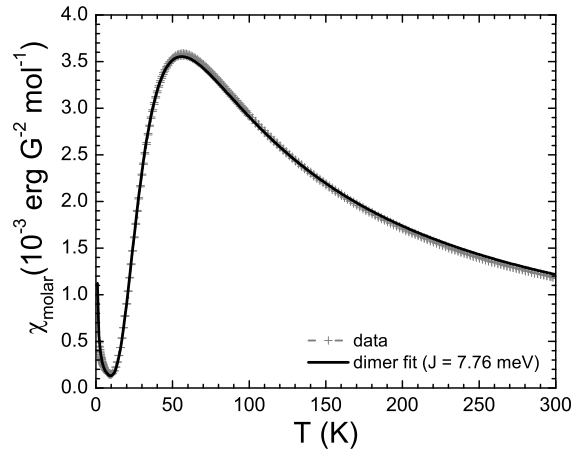}
\vskip -0.2cm
\caption{A fit of the dimer susceptibility formula of Eq.(\ref{chi_dimer_eq})
to the measured susceptibility of VO(HPO$_4$)$\cdot$0.5H$_2$O.
A defect term was also included.}
\label{chi_vohpo}
\end{figure}

As an example of the application of
the dimer susceptibility of Eq.(\ref{chi_dimer_eq})
(known as the Bleaney-Bowers formula \cite{Ble52}),
in Fig.\ref{chi_vohpo} we show a fit
to the susceptibility of the spin dimer
VO(HPO$_4$)$\cdot$0.5H$_2$O \cite{chi_vohpo}.
(The molar susceptibility shown is related
to the single dimer susceptibility of Eq.(\ref{chi_dimer_eq})
by $\chi_{molar} = N_A / 2 \cdot \chi$.)
The parameters of the fit are $g = 2.05$ and J $ = 7.76$~meV
(consistent with the results of inelastic neutron scattering \cite{Ten97}).
A 1/T defect contribution was also included in the fit.

\begin{figure}[ht]
\includegraphics[width=3.4in]{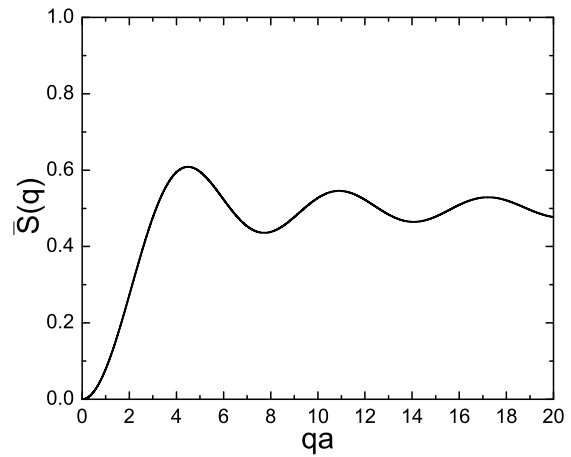}
\vskip -0.2cm
\caption{The powder average, unpolarized neutron structure factor $\bar S(q)$
for a spin dimer with pointlike magnetic ions.}
\label{dimerint}
\end{figure}

Finally we evaluate the inelastic neutron scattering intensities, which are
given by the structure factors of Eqs.(\ref{Strunpoldefn},\ref{Strpowavg}).
(A complete set of inelastic neutron scattering transitions for all the
spin systems we consider in this work
is given in Table~\ref{INSE}; typically we will
only evaluate the structure factors for the ground state of the
antiferromagnetic system.)
We evaluate Eq.(\ref{Strunpoldefn}) for the dimer using the
energy eigenvectors
$|\Psi_1(m)\rangle$
and
$|\Psi_0\rangle$
of Eqs.(\ref{dimervec1},\ref{dimervec0}). This gives
\be
S(\vec q\, ) = \frac{1}{2}
\big( 1 - \cos(\vec q \cdot \vec a\, ) \big)
\label{Sab_dimer}
\ee
where $\vec a = \vec x_1 - \vec x_2 = \vec x_{12}$
is a spatial vector that coincides with the dimer.
Evidently there should be no excitation of the dimer spin-triplet
state when the neutron momentum transfer $\vec q$ is perpendicular
to the dimer axis $\hat a$.

In scattering from powder samples one measures the powder average
${\bar S}(q)$ of the structure
factor, defined by Eq.(\ref{Strpowavg}). For the dimer this is
\be
{\bar S}(q) = \int \frac{d\Omega_{\hat q}}{4\pi}\,
\frac{1}{2}\, \big( 1 - \cos(\vec q \cdot \vec a\, )\big) =
\frac{1}{2}\,
\big( 1 - j_0(qa) \big)
\label{dimerinteq}
\ee
where $j_0(x) = \sin(x)/x$ is a spherical Bessel function.
This result is shown in Fig.\ref{dimerint} for pointlike magnetic ions
(F$(\vec q\, ) = 1$).
The location of the first maximum, at $q\approx 4.493\, a^{-1}$,
provides a convenient estimate of the
separation between the interacting ions in the dimer.
Of course in real materials the incorporation of ionic form factors
will reduce the location of this maximum.

Experimental studies of real magnetic materials typically proceed
by establishing the approximate magnetic parameters of a model
Hamiltonian through a fit to the susceptibility. Given a model
Hamiltonian, one can predict the inelastic neutron scattering
structure factor, which is then compared to experiment. (Ideally
this is done on single crystal samples, but frequently only
powder samples are available.) Unlike the bulk susceptibility,
the inelastic neutron scattering structure factor allows a sensitive
and microscopic test of the assumed magnetic Hamiltonian, since it is
determined by the relative positions of the interacting
magnetic ions. The spin-dimer material VO(DPO$_4$)$\cdot$0.5D$_2$O
provides a recent illustration of the use of inelastic neutron scattering in
identifying magnetic interaction pathways; the susceptibility
data of Johnson {\it et al.} \cite{Joh84}
was well known to give an excellent fit to the dimer formula
Eq.(\ref{chi_dimer_eq}), however the separation
of the interacting V-V pair inferred from
inelastic neutron scattering data \cite{Ten97}
using Eq.(\ref{dimerinteq})
showed that the interacting V-V pair had been misidentified
in the literature.

\subsection{Trimers}

We will consider the most general case of a spin trimer
with Heisenberg magnetic interactions. It is useful to
present the results as special cases with decreasing symmetry, since the
formulas are simpler in the more symmetric cases, and examples of
both symmetric and isosceles trimers are known in the literature.

\begin{figure*}[ht]
\includegraphics[height=3in,width=2in]{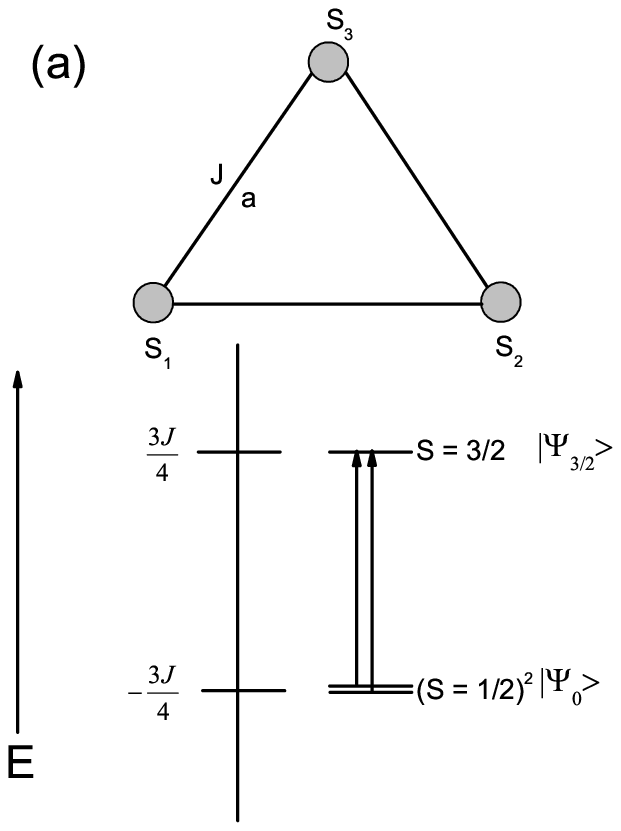}
\includegraphics[height=3in,width=2in]{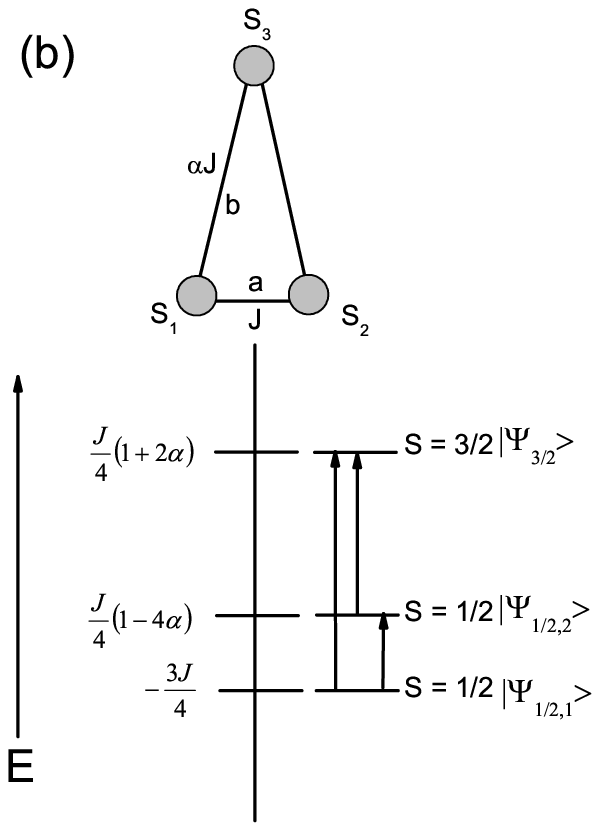}
\includegraphics[height=3in,width=2.25in]{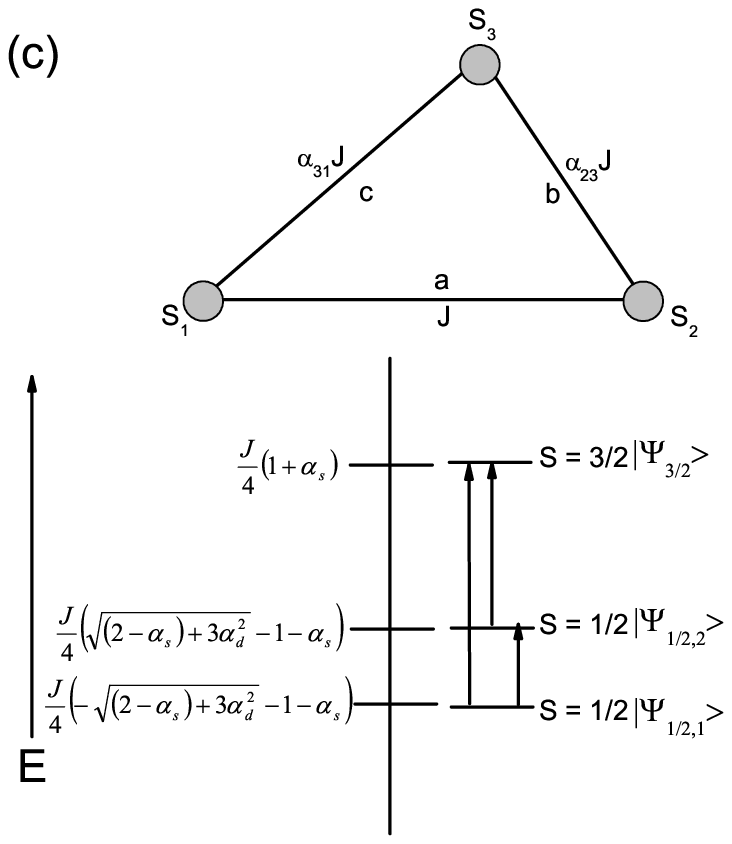}
\caption{Geometries and energy levels of (a) symmetric,
(b) isosceles and (c) general trimer systems. These systems have
one S$_{tot} = \frac{3}{2}$ multiplet and two
S$_{tot} = \frac{1}{2}$ multiplets.}
\label{123trimerdia}
\end{figure*}

\subsubsection{Symmetric Trimer}

The completely symmetric, equilateral trimer has
equal magnetic couplings and bond lengths between all three
pairs of spins.
The Hamiltonian for this model is
\be
{\cal H}
=
{\rm J}\,
\Big(
\vec{\S}_1 \cdot \vec{\S}_2 + \vec{\S}_2 \cdot \vec{\S}_3 +
\vec{\S}_3 \cdot \vec{\S}_1
\Big)\ .
\label{EtrimerH}
\ee
Since this Hamiltonian is invariant under any permutation
of the three spin labels,
it has a discrete S$_3$ symmetry in addition to
the magnetic rotational symmetry. In an S$_z$-diagonal basis
\vskip -0.5cm
\be
\left[
\begin{array}{l}
|\!\upa\upa\upa\,\rangle
\\
|\!\upa\upa\dna\,\rangle
\\
|\!\upa\dna\upa\,\rangle
\\
|\!\dna\upa\upa\,\rangle
\\
|\!\upa\dna\dna\,\rangle
\\
|\!\dna\upa\dna\,\rangle
\\
|\!\dna\dna\upa\,\rangle
\\
|\!\dna\dna\dna\,\rangle
\\
\end{array}
\right] 
\ee
the Hamiltonian matrix is
\begin{eqnarray}
{\rm J}
\left[
\begin{array}{cccccccc}
3/4 &       &        &       &  & & &          \\
     &  -1/4  &  \-1/2  & \-1/2  & & & &       \\
     & \-1/2  &   -1/4  & \-1/2  & & & &       \\
     & \-1/2  &  \-1/2  &  -1/4  & & & &       \\
     & & & &      -1/4  & \-1/2  & \-1/2     & \\
     & & & &     \-1/2  &  -1/4  & \-1/2     & \\
     & & & &     \-1/2  & \-1/2  &  -1/4     & \\
     &       &        &       &  & & &   3/4   \\
\end{array}
\right]_{\ .}
\nonumber
\end{eqnarray}
\be
\label{EtrimerHmat}
\ee
This matrix is block diagonal within subspaces of definite S$_{z\, tot}$,
as expected for a rotationally invariant Hamiltonian.
The energy levels of the symmetric trimer are shown in
Fig.\ref{123trimerdia}a. For the J~$ > 0$ (antiferromagnetic) case
the ground state is a quadruplet
(the two S$_{tot} = 1/2$ multiplets are degenerate), and there is
an energy gap of $\frac{3}{2}$J to the S$_{tot} = 3/2$ excited state.
Representative symmetric trimer energy eigenstates (those with maximum
S$_{z\, tot}$) are given in Table~\ref{wavefunctions1}.
Since the two S$_{tot} = 1/2$ levels are degenerate, there is no unique
ground state for this system; we use the Jacobi
$|\rho\rangle = |(12)_A\rangle$ and $|\lambda\rangle = |(12)_S\rangle$
three-body basis states of definite $(12)$-exchange symmetry as our
two independent basis vectors.

We may determine the specific heat and magnetic
susceptibility of the symmetric trimer from these
energy levels, using Eqs.(\ref{C},\ref{chi}).
The results are

\be
\frac{C}{k_B} =
\frac{9}{4}(\beta\J )^2
\frac
{e^{- \frac{3}{2} \beta \J}}
{(1 + e^{- \frac{3}{2} \beta \J} )^2 }
\label{C_strimer}
\ee
and
\be
\frac{\chi}{(g\mu_B)^2/{\J}} =
\frac{1}{4}\, \beta \J \,
\frac
{( 1 + 5 e^{-\frac{3}{2} \beta \J } )}
{( 1 + e^{-\frac{3}{2}\beta \J} ) }
\ .
\label{chi_strimer}
\ee
It is notable that the integral of this specific heat gives an entropy of
\be
S = \int_0^{\infty}\!\! C \; \frac{d\beta}{\beta} = k_B \ln(2)
\label{Entropy_strimer}
\ee
which is only half as large as the entropy of the dimer, despite the
larger trimer Hilbert space,
${\cal N} = 2^3 = 8$. The lower entropy is due to the
fourfold degenerate ground state
of this highly frustrated system;
\be
S = k_B \ln({\cal N}/{\cal N}_0)
  = k_B \ln(2^3/4) = k_B \ln(2)\ .
\label{Entropy_strimer2}
\ee

\begin{figure} [h]
\includegraphics[width=3.75in]{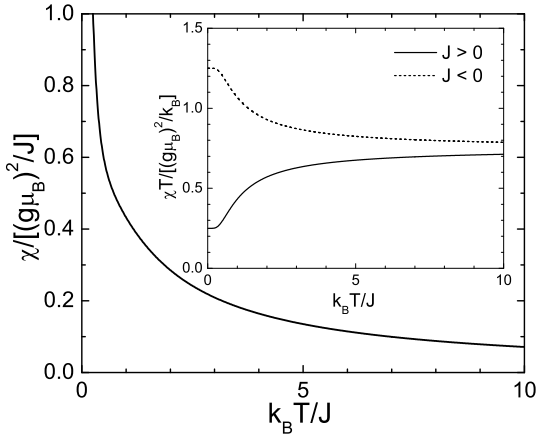}
\vskip -1.0cm
\caption{The magnetic susceptibility of a symmetric trimer.
The inset shows
$\chi T$ versus $T$ for
ferromagnetic (dashed) and antiferromagnetic (solid)
couplings.}
\label{trimersus}
\end{figure}

The susceptibility of the symmetric trimer, Eq.\ref{chi_strimer},
agrees with Eq.2 of Veit {\it et al.} \cite{Vei86}
(after specializing to a single $g$-factor and a change of variables).
This result is shown in Fig.\ref{trimersus}. Note that $\chi(T)$
diverges as we approach zero temperature,
since the ground state has nonzero spin. This divergence
is present independent of whether the intrinsic
spin-spin coupling J is antiferromagnetic (as we normally
assume) or ferromagnetic, since both cases have
ground states of nonzero spin.
A more detailed comparison of the susceptibility suffices to distinguish
these; see the inset of Fig.\ref{trimersus}, which shows
$\chi$T versus T for both cases.
At high temperatures the spin-spin coupling J is unimportant,
and both results approach the same Curie's law limit.

Next we consider the neutron scattering structure factors for the
symmetric trimer.
Since this system has two degenerate S$_{tot}$ = 1/2 ground
states and a single S$_{tot}$ = 3/2 excitation, there are two
distinct exclusive inelastic neutron structure
factors
but only a single transition energy,
${\rm E}_1 - {\rm E}_0 = \frac{3}{2}\,$J.
We have chosen $|\rho\rangle$ and $|\lambda\rangle$
basis states for our orthogonal S$_{tot}$ = 1/2 eigenstates,
and will give neutron
structure factors for each of these. The same structure factors
follow for the isosceles trimer, although in that case the two
S$_{tot}$ = 1/2 states are nondegenerate.

The structure factors for
excitation of the S$_{tot}$ = 3/2
$|\sigma\rangle$ level (using Eq.(\ref{Strunpoldefn})) are given by
\be
S^{(\rho\to\sigma)}(\vec q\, ) =
\frac{1}{3}\big( 1 - \cos(\vec q \cdot \vec x_{12}) \big) \ ,
\label{strimer_SI}
\ee
\bd
S^{(\lambda\to\sigma)}(\vec q\, ) = \hskip 3cm
\ed
\be
\frac{1}{3}\,
\big(
  1
+ \frac{1}{3} \cos(\vec q \cdot \vec x_{12})
- \frac{2}{3} \cos(\vec q \cdot \vec x_{13})
- \frac{2}{3} \cos(\vec q \cdot \vec x_{23})
\big)\ .
\label{strimer_SII}
\ee

These results may be understood in terms of
the different natures of the
$|\rho\rangle$ and $|\lambda\rangle$ initial states.
In the $|\rho\rangle$
ground state the (12)-dimer is in a pure S$_{(12)} = 0$
state, which must be excited to S$_{(12)} = 1$ to couple to the
$|\sigma\rangle$ excited state. The $|\rho\rangle \to |\sigma\rangle$
excitation problem is thus
identical to the dimer problem,
to within an overall constant.
It follows that $S^{(\rho\to\sigma)}(\vec q\, )$
is proportional to the dimer structure
factor of Eq.(\ref{Sab_dimer}).
In contrast, in the $|\lambda\rangle$ initial state the
(12)-dimer is pure S$_{(12)} = 1$ and the (23)- and
(31)-dimers have amplitudes to be in both spin 0 and 1,
so there are contributions to $S^{(\lambda\to\sigma)}$
due to the excitation of each of the three dimer subsystems.

As $S^{(\rho\to\sigma)}$ and $S^{(\lambda\to\sigma)}$ differ
considerably for moderate $qa$ it will certainly
be possible to distinguish between $|\rho\rangle$ and
$|\lambda\rangle$ states from their single crystal structure
factors. The powder averages however are identical, and cannot
be distinguished experimentally;

\be
{\bar S}^{(\rho\to\sigma)}(q) =
{\bar S}^{(\lambda\to\sigma)}(q) =
\frac{1}{3}\big( 1 - j_0(q a) \big) \ .
\label{Strimerpowavg}
\ee

These powder structure factors are identical because of
the identical dimer lengths,
$r_{12} = r_{23} = r_{31} = a$, so the powder average of each
cosine in Eqs.(\ref{strimer_SI},\ref{strimer_SII})
gives the same $j_0(q a)$ Bessel function. The dependence
$(1 - j_0(q a))$ follows from the requirement that $\bar S(0) = 0$.

As we shall discuss in the next section,
an isosceles trimer would be a more favorable system
for the identification of $|\rho\rangle$ and $|\lambda\rangle$ initial states
in inelastic neutron scattering; these levels are nondegenerate
in the isosceles system, and the
$|\rho\rangle\to|\sigma\rangle$ and $|\lambda\rangle\to|\sigma\rangle$
powder average structure factors
are no longer equal, due to the different leg lengths.

\subsubsection{Isosceles Trimer}

The isosceles spin trimer, Fig.~\ref{123trimerdia}b,
has two equal magnetic interactions and bond lengths.
The Hamiltonian is given by
\be
{\cal H}
= {\rm J}
\Big(
\vec{\S}_1 \cdot \vec{\S}_2 + \alpha
\big( \vec{\S}_2 \cdot \vec{\S}_3 + \vec{\S}_3 \cdot \vec{\S}_1\big)
\Big)\ .
\label{isotrimerH}
\ee
To find the energy eigenvalues of this Hamiltonian
it suffices to consider the S$_{z\, tot} = +1/2$
sector, since the S$_{tot} = 1/2$ and $3/2$ multiplets
both have S$_{z\, tot} = +1/2$ members.
The remaining symmetry of this problem suggests that we use the
three $\{ |{\rm S}_{tot}, +1/2\rangle \} $
energy eigenstates of the symmetric trimer as our basis,
\be
\left[
\begin{array}{l}
|\sigma(3/2,+1/2)\rangle =
\sqrt{\frac{1}{3}}\
(|\!\upa\upa\dna\,\rangle +
 |\!\upa\dna\upa\,\rangle +
 |\!\dna\upa\upa\,\rangle )
\\
|\lambda(1/2,+1/2)\rangle =
\sqrt{\frac{1}{6}}\
(|\!\upa\dna\upa\,\rangle + |\!\dna\upa\upa\,\rangle
- 2 |\!\upa\upa\dna\,\rangle )
\\
|\rho(1/2,+1/2)\rangle =
\sqrt{\frac{1}{2}}\
(|\!\upa\dna\upa\,\rangle - |\!\dna\upa\upa\,\rangle )
\\
\end{array}
\right]_{\ .}
\label{trimerbasis}
\ee
The Hamiltonian is necessarily diagonal on this basis,
since these three basis states have different values of the
conserved quantities
S$_{tot}$ and $(12)$-exchange symmetry. The result is
\begin{eqnarray}
{\cal H}
=
\frac{1}{4} \,
{\rm J}
\left[
\begin{array}{ccc}
1 + 2\,\alpha &    & \\
 & 1 -  4\,\alpha   & \\
       &   & - 3
\end{array}
\right]_{\ .}
\label{isotrimerHmatrix}
\end{eqnarray}
The two
S$_{tot} = \frac{1}{2}$ levels are split
as a result of the reduced symmetry of the isosceles
trimer; the full
S$_3$ symmetry of the symmetric trimer has been
reduced to S$_2$ ($(12)$-exchange symmetry), and
since S$_2$ is Abelian
no degeneracies follow from this symmetry.

The specific heat and susceptibility of the isosceles trimer,
which follow from the energy levels of Eq.(\ref{isotrimerHmatrix})
and the formulas Eqs.(\ref{C},\ref{chi}), are given in
Tables~\ref{C_table} and \ref{chi_table}.
The susceptibility agrees with the earlier result
of Veit {\it et al.} \cite{Vei86}.
Note that one recovers the symmetric trimer result in the limit
$\alpha = 1$.

We have confirmed by numerical integration of the
rather complicated isosceles trimer specific heat formula
given in Table~\ref{C_table} that the entropy
of the isosceles trimer satisfies
\be
\frac{S}{k_B} =
\int_0^{\infty}\!\! \frac{C}{k_B} \; \frac{d\beta}{\beta} =
\begin{cases}
2\ln(2) &           \text{$\alpha \neq 1 $}\\
\phantom{2}\ln(2) & \text{$\alpha =  1 $\ ,}
\end{cases}
\ee
as expected from Eq.(\ref{Entropy_general})
for an eight-dimensional Hilbert space which
has a fourfold-degenerate ground state for $\alpha = 1 $,
and a twofold-degenerate ground state otherwise.

Although the magnetic contribution to the specific
heat is usually masked by much larger phonon contributions,
we note in passing that one may separate magnetic contributions 
experimentally by subtracting the specific heats in zero and nonzero 
magnetic fields. This approach was 
used recently by Luban {\it et al.} \cite{Lub02} to study an 
S=1/2 V$^{4+}$ vanadium trimer,
(CN$_3$H$_6$)$_4$Na$_2$[H$_4$V$_6$O$_8$(PO$_4$)$_4$
((OCH$_2$)$_3$CCH$_2$OH)$_2$]$\cdot 14$H$_2$O
(their material~{\bf 1}),
which appears to be an accurate realization of the isosceles 
Heisenberg trimer. 

As we found for the symmetric trimer, the susceptibility
of the isosceles trimer also diverges as T approaches zero,
since the system has a magnetized ground state.
The rate of divergence with T can again be used to distinguish
between ferromagnetic and antiferromagnetic couplings
(which have S$_{tot}=3/2$ and S$_{tot}=1/2$ ground states respectively),
as shown in Fig.\ref{trimersus} for the symmetric trimer. 
This behavior is evident in the susceptibility of
material~{\bf 1} of Luban {\it et al.} \cite{Lub02}; 
in Fig.3 of this reference one can see that $\chi$T for this material 
clearly follows the lower trimer curve, confirming that it is 
accurately described by the antiferromagnetic 
isosceles trimer model (with an S$_{tot} = 1/2$ ground state).

There are three inelastic transitions excited by
neutron scattering from an isosceles spin trimer,
$|\rho\rangle \to |\sigma\rangle$,
$|\lambda\rangle \to |\sigma\rangle$ and
$|\rho\rangle \to |\lambda\rangle$. The first two were considered in the discussion
of the symmetric trimer, and the results for the isosceles trimer are
identical (except that the $\Delta E$ values of the transitions differ).
The $|\rho\rangle \to |\lambda\rangle$ transition was not
considered previously because
these states are degenerate in the symmetric trimer. The result we
find for the structure factor of this transition is
\be
S^{(\rho\to\lambda)}(\vec q\, ) =
\frac{1}{6}\Big( 1 - \cos(\vec q \cdot \vec x_{12}) \Big) \ .
\label{itrimer_SIII}
\ee
This has the same form as the dimer and
$|\rho\rangle \to |\sigma\rangle$ structure
factors because it also involves the excitation of the
S$_{(12)}$ = 0 (12)-dimer
to an S$_{(12)}$ = 1 state. It can evidently be distinguished from
the $|\rho\rangle \to |\sigma\rangle$
transition by the overall intensity, but not by the functional
dependence on $q$.

\begin{figure}[ht]
\includegraphics[width=3.4in]{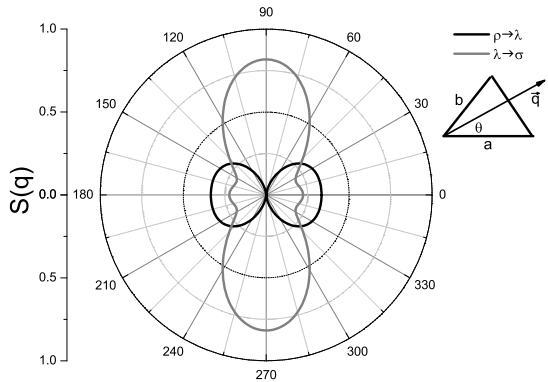}
\caption{The unpolarized structure factors
$\{S(\vec q)\} $ (proportional to the angular scattering intensities)
predicted for inelastic neutron scattering from the
$|\lambda\rangle$ ground state of a single crystal of an isosceles trimer
material (see text).}
\label{isoSabxtalfig}
\end{figure}

To illustrate these single crystal
structure factors, in Fig.\ref{isoSabxtalfig}
we show the two ground state
structure factors for the previously cited V$_6$ isosceles
trimer
(material~{\bf 1} of Luban {\it et al.} \cite{Lub02}).
The parameters are $a = 3.22$~\AA\; and $b = 3.364$~\AA. We show
the predictions of Eqs.(\ref{strimer_SII},\ref{itrimer_SIII}) for
in-plane scattering, with momentum transfer $q = \pi/a$.
Since this material has two strong bonds and a weak dimer
$(\alpha \approx 9)$ \cite{Lub02},
$|\lambda\rangle$ should be the ground state, and the
$|\lambda\rangle \to |\rho \rangle $
and
$|\lambda\rangle \to |\sigma \rangle $ transitions shown
in the figure should both be observable (These are expected at 4.2~meV and
7.0~meV respectively, given the parameters of Luban {\it et al}.)
The very different angular distributions predicted for the scattered
neutrons show that it should be straightforward to distinguish between
these transitions in an inelastic neutron scattering experiment,
given a single crystal of this or a similar trimer material.

The powder average eliminates much of the difference between these
neutron scattering transitions, although it still should be possible
to distinguish them experimentally.
On carrying out the powder average we find

\begin{eqnarray}
\begin{array}{ccc}
&{\bar S}^{(\rho\to\sigma)}(q) =
&\frac{1}{3}\big( 1 - j_0(q a) \big)
\phantom{\ .}
\\
\\
&{\bar S}^{(\lambda\to\sigma)}(q) =
&\frac{1}{3}\big( 1 + \frac{1}{3}j_0(q a) - \frac{4}{3} j_0(q b) \big)
\\
\\
&{\bar S}^{(\rho\to\lambda)}(q) =
&\frac{1}{6}\big( 1 - j_0(q a) \big)
\phantom{\ .}
\ .
\\
\end{array}
\label{isotrimerpowavg}
\end{eqnarray}

\begin{figure}[t]
\includegraphics[width=3.4in]{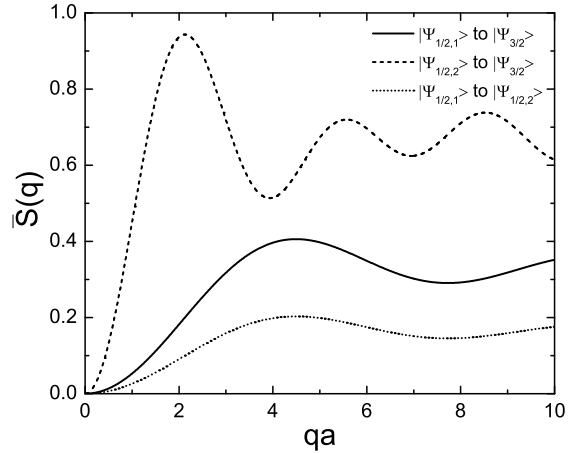}
\caption{The powder average inelastic neutron structure factor $\bar S(q)$
for the three allowed transitions of an isosceles trimer,
with $b/a = 2$.}
\label{isotrimerpaie2}
\end{figure}

In the symmetric limit $b/a = 1$ these transitions
are proportional to the same function,
$1 - j_0(q a)$; at best it may be possible to
distinguish the $|\rho\rangle\to|\lambda\rangle$
transition from the others
through their relative intensities.
However for significantly different leg lengths
the $|\lambda\rangle\to |\sigma\rangle$
powder average structure factor
may differ enough from the $1 - j_0(q a)$ of the
$|\rho\rangle\to |\sigma\rangle$
and
$|\rho\rangle\to |\lambda\rangle$
transitions to distinguish them.
As an example, in Fig.\ref{isotrimerpaie2}
we show the powder structure factors of Eq.(\ref{isotrimerpowavg})
for the three transitions, for an elongated triangle with $b/a = 2$.
(These results are independent of the
magnetic coupling ratio $\alpha$.)
As there is considerable variation in form and magnitude between these
powder structure factors,
it should be possible to distinguish them experimentally
in similar isosceles trimer materials.
If more than one transition is clearly observed, it may also be
useful to compare structure factor ratios, to eliminate
the effect of ionic form factors.

\subsubsection{General Trimer}

The general trimer of Fig.\ref{123trimerdia}c
has three different magnetic couplings and
ion pair separations, and is described by the
Hamiltonian
\be
{\cal H}
= {\rm J}
\Big(
\vec{\S}_1 \cdot \vec{\S}_2 +
\alpha_{23}\, \vec{\S}_2 \cdot \vec{\S}_3 +
\alpha_{31}\, \vec{\S}_3 \cdot \vec{\S}_1
\Big)\ .
\label{RtrimerH}
\ee
This system is also discussed by Qiu {\it et al.} \cite{Qiu02}
in the context of La$_4$Cu$_3$MoO$_{12}$, which they model as
a two dimensional coupled array of S=1/2 trimers.

We may again determine all the trimer
energy eigenvalues by specializing to the
S$_{z\, tot} = +1/2$ sector and using the symmetric trimer basis
of Eq.(\ref{trimerbasis}), which gives the Hamiltonian matrix
\begin{eqnarray}
{\cal H}
= \frac{1}{4}\,{\rm J}
\left[
\begin{array}{ccc}
1 + \alpha_s
&
&
\\

&
1 - 2\alpha_s
&
\sqrt{3}\, \alpha_d
\\

&
\sqrt{3}\, \alpha_d
&
-3
\end{array}
\right]
\label{rtrimerHmatrix}
\end{eqnarray}
where
$\alpha_s = \alpha_{31} + \alpha_{23}$
and
$\alpha_d = \alpha_{31} - \alpha_{23}$.
The $|\sigma\rangle$ basis states again must be energy eigenstates,
since they are the only S$_{tot}=3/2$ states in the Hilbert space.
They have energies of
\be
{\rm E}_{\frac{3}{2}} =
\frac{ 1 +\alpha_s }{4}\, {\rm J}\, \ .
\label{E_gentrimer1}
\ee
The
S$_{tot}=1/2$ basis states
$|\rho\,\rangle$
and
$|\lambda\rangle$
mix in this problem, since the general trimer Hamiltonian
with $\alpha_{23} \neq \alpha_{31}$ breaks
(12)-exchange symmetry. The resulting energies are
\be
{\rm E}_{\frac{1}{2}, \, \{ {1\atop 2}\} } =
-
\frac{ ( 1 + \alpha_s  \pm
\sqrt{ ( 2 - \alpha_s )^2 + 3\,\alpha_d^2\, }\, ) }{4}
\; {\rm J} \ .
\label{E_gentrimer2}
\ee

The specific heat and susceptibility of the general trimer
follow from the energy levels of
Eqs.(\ref{E_gentrimer1},\ref{E_gentrimer2})
and the formulas Eqs.(\ref{C},\ref{chi}). The resulting
expressions are given in Tables~\ref{C_table} and \ref{chi_table}.
One may confirm recovery of the isosceles and symmetric trimer
results as special cases of these results. We have also
confirmed by numerical integration that the rather lengthly
general trimer specific heat formula given
in Table~\ref{C_table} leads to an entropy of
$S = k_B\, 2 \ln(2)$, provided that at least one of the
parameters $\alpha_{23}$ and $\alpha_{31}$ differs from unity.

The neutron scattering structure factors for the general trimer
involve coherent superpositions of the previously derived
$|\rho\rangle$ and $|\lambda\rangle$ excitation functions, since the
energy eigenstates are superpositions of these basis states.
The S$_{tot}=1/2$ energy eigenstates of
Eq.(\ref{rtrimerHmatrix}) are explicitly
\be
|\Psi_{\frac{1}{2},1}\rangle =
-
\sin(\theta) |\lambda\rangle
+
\cos(\theta) |\rho\rangle
\label{gtstateh1}
\ee
and
\be
|\Psi_{\frac{1}{2},2}\rangle =
+
\cos(\theta) |\lambda\rangle
+
\sin(\theta) |\rho\rangle \ ,
\label{gtstateh2}
\ee
where the mixing angle $\theta$ satisfies
\be
\tan(\theta) = \frac{x}{1+\sqrt{1+x^2}}
\ee
with $x = \sqrt{3}\, \alpha_d/(2-\alpha_s)$.
The S$_{tot}=3/2$ energy eigenstate is, as for all the
trimers we have considered,
\be
|\Psi_{\frac{3}{2}}\rangle = |\sigma\rangle \ .
\ee
The structure factor for the
transition from the
S$_{tot}=1/2$ state
$|\Psi_{\frac{1}{2},1}\rangle $
to the S$_{tot}=3/2$ state is given by
\bd
S^{(\Psi_{\frac{1}{2},1}\to\Psi_{\frac{3}{2}})}(\vec q\, ) =
\frac{1}{3}
\bigg(
\big(1 - \frac{1}{3}(C_{12} + C_{23} + C_{31})\big)
\ed
\be
+ \frac{1}{3}\, {\cal C}_2\,
\big(\, C_{31} + C_{23} - 2\, C_{12}\big)
+ \frac{1}{\sqrt{3}}\, {\cal S}_2\,  \big(C_{31} - C_{23}\big)
\bigg)
\label{gtrimer_SI}
\ee
where
$C_{ij} = \cos(\vec q \cdot \vec x_{ij})$,
${\cal C}_2 = \cos(2\theta)$
and
${\cal S}_2 = \sin(2\theta)$.
The structure factor for the second transition,
$\Psi_{\frac{1}{2},2}\to\Psi_{\frac{3}{2}}$,
follows from Eq.(\ref{gtrimer_SI}) on changing the overall signs of
the ${\cal C}_2$ and ${\cal S}_2 $ terms.
The third transition, between the two S$_{tot}=1/2$ states,
has the structure factor
\bd
S^{(\Psi_{\frac{1}{2},1}\to\Psi_{\frac{1}{2},2})}(\vec q\, ) =
\frac{1}{6}
\bigg(
\big(1 - \frac{1}{3}(C_{12} + C_{23} + C_{31})\big)
\ed
\be
+ \frac{1}{3}\, {\cal C}_4\, \big(C_{31} + C_{23} - 2\, C_{12}\big)
+
\frac{1}{\sqrt{3}}\, {\cal S}_4\, \big( C_{31} - C_{23}\big)
\bigg)\ ,
\label{gtrimer_SIII}
\ee
where the new quantities are
${\cal C}_4 = \cos(4\theta)$
and
${\cal S}_4 = \sin(4\theta)$.
One may confirm that the previously derived symmetric and
isosceles trimer structure factors of
Eqs.(\ref{strimer_SI},\ref{strimer_SII},\ref{itrimer_SIII})
follow from these
general trimer results in the limit $\theta\to 0$.

The powder averages of these general trimer
unpolarized structure factors may also
be evaluated; the result for the transition
$\Psi_{\frac{1}{2},1}\to\Psi_{\frac{3}{2}}$ is
\bd
{\bar S}^{(\Psi_{\frac{1}{2},1}\to\Psi_{\frac{3}{2}})}(\vec q\, ) =
\frac{1}{3}
\bigg(
1
- \frac{(1 + 2\,{\cal C}_2)}{3}\, j_0(q r_{12})
\ed
\be
- \frac{(1 - {\cal C}_2 + \sqrt{3}\, {\cal S}_2 )}{3}\, j_0(q r_{23})
- \frac{(1 - {\cal C}_2 - \sqrt{3}\, {\cal S}_2) }{3}\, j_0(q r_{31})
\bigg) \ .
\label{gtrimerpowavg}
\ee
The powder average results for the two remaining transitions can be obtained
from Eq.(\ref{gtrimerpowavg}) by simple substitutions. To obtain
${\bar S}^{(\Psi_{\frac{1}{2},2}\to\Psi_{\frac{3}{2}})}$
simply change the overall signs of ${\cal C}_2$ and ${\cal S}_2$
in Eq.(\ref{gtrimerpowavg}), and to obtain
${\bar S}^{(\Psi_{\frac{1}{2},1}\to\Psi_{\frac{1}{2},2})}$,
divide Eq.(\ref{gtrimerpowavg}) by a factor of two and replace
${\cal C}_2$ and ${\cal S}_2$ by ${\cal C}_4$ and ${\cal S}_4$
respectively.

These results will be useful for the interpretation of
neutron scattering data on real materials.
One example of a candidate general trimer is the V$_6$
material~{\bf 2} of Luban {\it et al.} \cite{Lub02},
Na$_6$[H$_4$V$_6$O$_8$(PO$_4$)$_4$((OCH$_2$)$_3$ CCH$_2$OH)$_2$]$\cdot 18$H$_2$O.
This compound has
three distinct V-V separations between the S=1/2 V$^{4+}$ ions within each
vanadium trimer, $3.212$~\AA, $3.252$~\AA\ and $3.322$~\AA.

\subsection{Tetramers}

\begin{figure*}[ht]
\includegraphics[height=3in,width=2in]{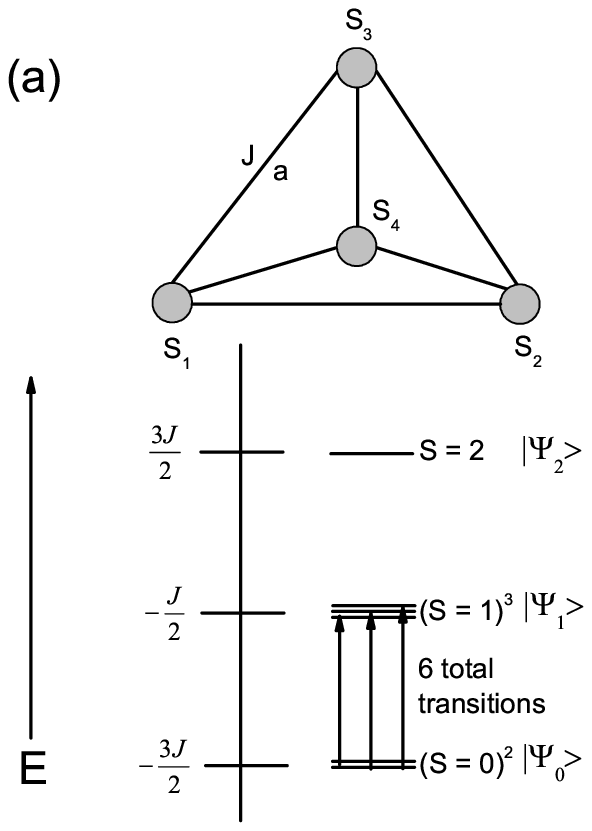}
\includegraphics[height=3in,width=2in]{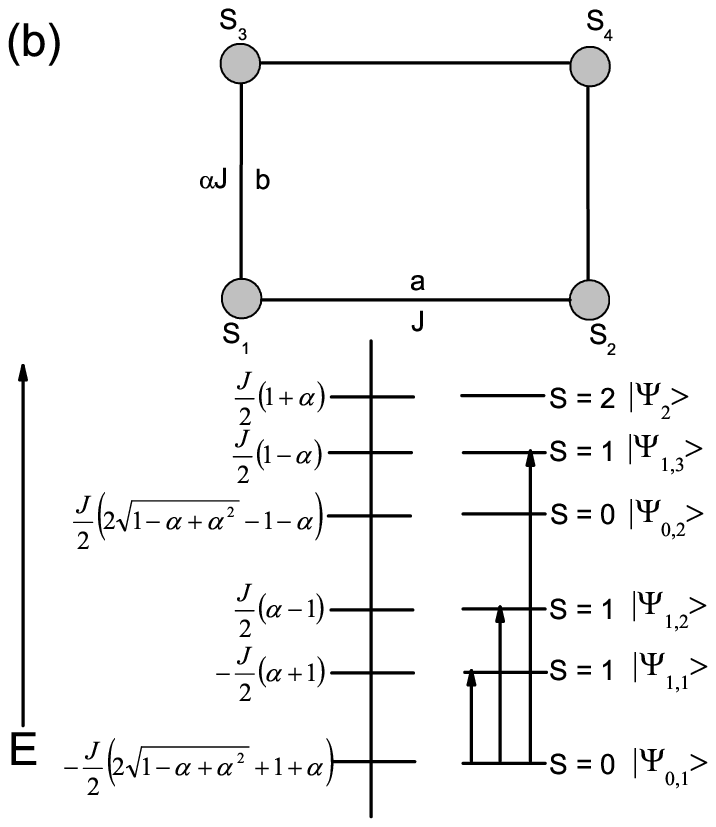}
\includegraphics[height=3in,width=2in]{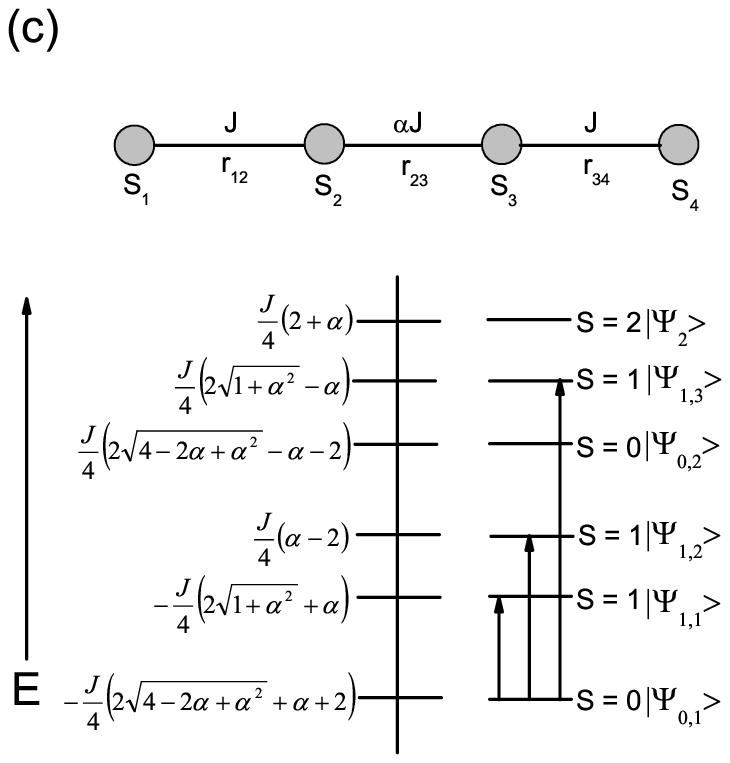}
\caption{Energy level diagrams for (a) the tetrahedron,
(b) the rectangular tetramer, and (c) the linear (dimer-pair) tetramer.}
\label{123tetramerdia}
\end{figure*}

We will consider three S = 1/2 tetramer spin clusters of
decreasing symmetry, the regular tetrahedron, the rectangular tetramer,
and the linear (dimer-pair) tetramer. Our definitions for the magnetic
couplings and geometry of these systems are shown in
Fig.\ref{123tetramerdia}. As with the dimer and trimer systems we will
give results for the partition function, specific heat, magnetic
susceptibility and
neutron inelastic scattering structure factors, the latter for
both single crystal and powder average cases.

\subsubsection{Tetrahedron}

This system has four S = 1/2 ions at the vertices of a
regular tetrahedron, with Heisenberg interactions of strength J
between each pair of ions (see Fig.\ref{123tetramerdia}a).
The Hamiltonian of this system is given by
\be
{\cal H} =
{\rm J} \sum_{i,j=1\atop i<j }^4 {\vec \S}_i \cdot {\vec \S}_j \ .
\label{Htetrahedron}
\ee
The invariance of this Hamiltonian under permutation of any site labels
implies an S$_4$ symmetry, in addition to the spin rotation symmetry
SU(2). Since the group S$_4$ is non-Abelian and has
irreducible representations of dimensionality
${\underbar 1}$,
${\underbar 2}$
and
${\underbar 3}$,
we anticipate that one may find twofold and threefold degeneracies in
the spectrum of tetrahedron energy eigenstates. We will see that this
is indeed the case.

As with the dimer and symmetric trimer we may determine the energy
eigenvalues of this system by simply squaring the total spin operator
$ \vec \S_{tot} =  \sum_{i=1}^n \vec \S_i$, which gives for this case
\be
\E_{{\rm S}_{tot}} = \frac{1}{2}\,
{\rm J} \Big(\, {\S}_{tot} ({\S}_{tot} +1) - 3 \Big)
=
\begin{cases}
+\frac{3}{2}\, \J  & \text{S$_{tot}  = 2$}\\
-\frac{1}{2}\, \J  & \text{S$_{tot}  = 1$}\\
-\frac{3}{2}\, \J  & \text{S$_{tot}  = 0$.}
\end{cases}
\label{Etetrahedron}
\ee
The Clebsch-Gordon series of
Eq.(\ref{CGseries}) implies that these S$_{tot} = 1$ and S$_{tot} = 0$ energy
levels are respectively
threefold and twofold degenerate.

Given these energy levels,
the specific heat and susceptibility of the tetrahedron may then
be determined using Eqs.(\ref{C},\ref{chi}), with the results

\be
\frac{C}{k_B} =
\frac{9}{2}(\beta\J )^2 e^{- \beta \J}\;
\frac{(1 + 5 e^{-2\beta \J} + 10 e^{-3\beta \J})}
{\left(
1 + \frac{9}{2} e^{- \beta \J} + \frac{5}{2} e^{-3\beta \J}
\right)^2 }
\ee
and
\be
\frac{\chi}{(g\mu_B)^2/\J} =
3\beta\J\, e^{-\beta \J}
\frac{(1+\frac{5}{3} e^{-2\beta \J})}
{\left(
1 + \frac{9}{2} e^{- \beta \J} + \frac{5}{2} e^{-3\beta \J}
\right) }
\ .
\ee

\begin{figure}[ht]
\includegraphics[width=3.4in]{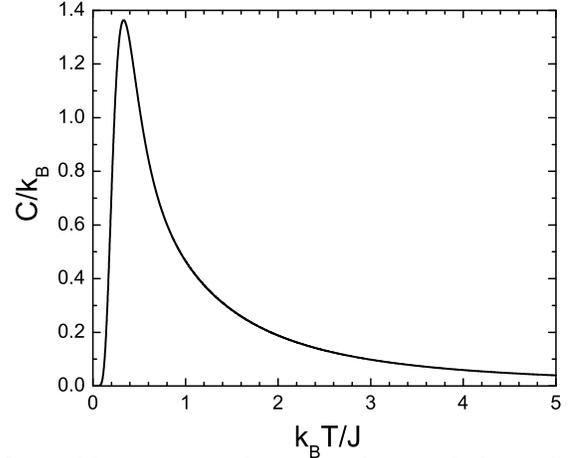}
\vskip -1.0cm
\caption{Magnetic contribution to the specific heat
of a regular tetrahedron.}
\label{cptetrahedron}
\end{figure}

\begin{figure}[ht]
\includegraphics[width=3.4in]{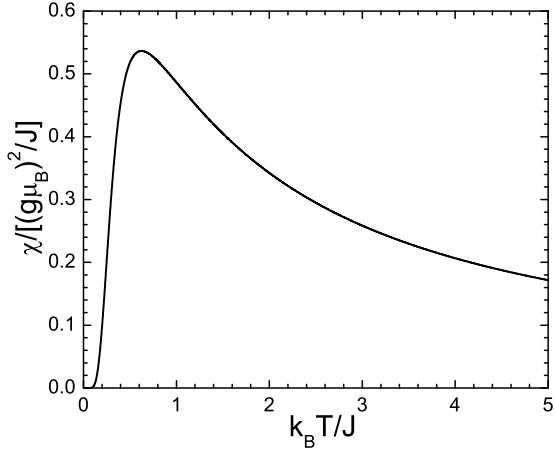}
\vskip -0.8cm
\caption{Susceptibility of a regular tetrahedron.}
\label{chitetrahedron}
\end{figure}

These quantities are shown in
Fig.\ref{cptetrahedron} and
Fig.\ref{chitetrahedron} respectively.
The specific heat of the tetrahedron gives an entropy of
$S = k_B\, 3\ln(2)$, as expected for a 16-dimensional Hilbert space
and a doubly-degenerate ground state.

Note that the
susceptibility is rather similar to that of the spin dimer, since the
tetrahedron also has an S$_{tot} = 0$ ground state and a gap
of J to the
magnetized S$_{tot} = 1$ excited states. (The fact that the
ground state is twofold degenerate does not affect this result,
since both are S$_{tot} = 0$ states and neither makes a contribution
to the susceptibility.)

Determination of the energy eigenvectors requires
diagonalization of the Hamiltonian on a specific basis.
Operating on our $|(12)(34)\rangle$ dimer basis of
Eqs.(\ref{Seq2tetramerbasis}-\ref{Seq0tetramerbasis})
with the tetrahedron Hamiltonian, Eq.(\ref{Htetrahedron}), we find
that the Hamiltonian matrix is already fully diagonal;
each of these basis states is an energy eigenvector of the
tetrahedron Hamiltonian.

In our discussion of neutron scattering structure factors of the
tetrahedron and the other spin tetramers considered in this paper,
we will specialize to S$_{\, tot} = 0$ initial states.
Structure factors for S$_{\, tot} > 0$ initial states, which
are of interest for systems with magnetized ground states
and at finite temperatures, and can be derived using similar methods.

The tetrahedron has two degenerate ground states, which we take to be
$|\Psi_{0,1}\rangle = |\rho\rho\rangle $
and
$|\Psi_{0,2}\rangle = |\sigma\sigma\rangle_0 $.
The three degenerate S$_{\, tot} = 1$ excited states, which can be reached
from the S$_{\, tot} = 0$ levels
using inelastic neutron scattering, are taken to be
$|\Psi_{1,1}\rangle = |(\rho\sigma)_S\rangle $,
$|\Psi_{1,2}\rangle = |(\rho\sigma)_A\rangle $ and
$|\Psi_{1,3}\rangle = |(\sigma\sigma)_S\rangle_1 $.
The choice of this specific set of initial and final states
is rather arbitrary; in a real material we would expect a spontaneous
distortion of the lattice, which would select nearly degenerate
energy eigenstates that need not be these specific basis states. However
these will suffice to illustrate the neutron scattering structure
factors expected for nearly tetrahedral systems.

The single crystal structure factors for all of these transitions
may be read directly from
Eqs.(\ref{tetramer_me2}-\ref{tetramer_me9}). For example, the transition
$|\Psi_{0,1}\rangle \to |\Psi_{1,1}\rangle $ is specified by the
matrix element of Eq.(\ref{tetramer_me2}); using the structure factor
definition in Eq.(\ref{Strunpoldefn}), we find
\bd
\hskip -4cm
S^{(\Psi_{0,1} \to \Psi_{1,1})}(\vec q\, ) = 1/2
\ed
\vskip -0.5cm
\be
-
\big(
 C_{12} - C_{13} + C_{14} + C_{23} - C_{24} + C_{34}
\big)/4
\label{Stetramer_01_to_11}
\ee
where as before $C_{ij} = cos(\vec q \cdot \vec x_{ij})$.
This characteristic angular distribution and its five partner
distributions could be used in an inelastic neutron
scattering experiment from a single crystal sample to characterize
the spin states of the individual
S$_{\, tot} = 0$ and S$_{\, tot} = 1$ levels. (Note however that
one specific transition, $|\Psi_{0,1}\rangle \to |\Psi_{1,3}\rangle $,
has a zero matrix element.)

The powder average structure factors for a tetrahedron are much less
characteristic. Since there is only a single ion pair separation, each
cosine in the single crystal structure factors such as
Eq.(\ref{Stetramer_01_to_11}) powder-averages to the same factor
of $j_0(qa)$. This gives a powder structure factor that is proportional
to $1-j_0(qa)$ for each transition, just as we found for the dimer and symmetric
tetramer; only the overall coefficients distinguish the different
transitions. These results are
\begin{eqnarray}
\begin{array}{ccc}
&{\bar S}^{(\Psi_{0,1} \to \Psi_{1,1})}(q) =
&\frac{1}{2}\big( 1 - j_0(q a) \big)
\phantom{\ .}
\\
\\
&{\bar S}^{(\Psi_{0,1} \to \Psi_{1,2})}(q) =
&\frac{1}{2}\big( 1 - j_0(q a) \big)
\phantom{\ .}
\\
\\
&{\bar S}^{(\Psi_{0,1} \to \Psi_{1,3})}(q) =
&0
\phantom{\ .}
\\
\\
&{\bar S}^{(\Psi_{0,2} \to \Psi_{1,1})}(q) =
&\frac{1}{6}\big( 1 - j_0(q a) \big)
\phantom{\ .}
\\
\\
&{\bar S}^{(\Psi_{0,2} \to \Psi_{1,2})}(q) =
&\frac{1}{6}\big( 1 - j_0(q a) \big)
\phantom{\ .}
\\
\\
&{\bar S}^{(\Psi_{0,2} \to \Psi_{1,3})}(q) =
&\frac{2}{3}\big( 1 - j_0(q a) \big)
\ .
\\
\end{array}
\label{tetrahedronpowavg}
\end{eqnarray}

A generalization of the tetrahedron problem
in which the Hamiltonian has couplings of strength
$\alpha$J between ions in different dimers
may also be of interest. This generalized Hamiltonian is
\bd
{\cal H}(\alpha) =
{\J }
\Big( \big(
\vec {\rm S}_{1}\cdot\vec {\rm S}_{2} +
\vec {\rm S}_{3}\cdot\vec {\rm S}_{4} \big)  \hskip 2cm
\ed
\be
\hskip 1cm +\; \alpha\,
\big(\,
\vec{\rm S}_{1}\cdot\vec{\rm S}_{3} +
\vec{\rm S}_{1}\cdot\vec{\rm S}_{4} +
\vec{\rm S}_{2}\cdot\vec{\rm S}_{3} +
\vec{\rm S}_{2}\cdot\vec{\rm S}_{4}\,
\big)\Big).
\label{Halphatetrahedron}
\ee
The dimer-pair basis of
Eqs.(\ref{Seq2tetramerbasis}-\ref{Seq0tetramerbasis})
is also diagonal under this Hamiltonian, with the eigenvalues
given below. (We have added an S$_{tot}$ subscript to all these
state vectors for clarity.)

\begin{eqnarray}
&&{\cal H}(\alpha)
|\sigma\sigma\rangle_{{\rm S}_{tot} = 2}
\hskip 0.05cm
=
\Big(\frac{1}{2} + \alpha\Big)\, |\sigma\sigma\rangle_2
\\
&&
{\cal H}(\alpha)
|\sigma\sigma\rangle_1
\hskip 0.75cm
=
\Big(\frac{1}{2} - \alpha\Big)\, |\sigma\sigma\rangle_1
\\
&&
{\cal H}(\alpha)
|(\rho\sigma)_S\rangle_1
\hskip 0.33cm
=
-\frac{1}{2}\,
|(\rho\sigma)_S\rangle_1
\\
&&
{\cal H}(\alpha)
|(\rho\sigma)_A\rangle_1
\hskip 0.30cm
=
-\frac{1}{2}\,
|(\rho\sigma)_A\rangle_1
\\
&&
{\cal H}(\alpha)
|\sigma\sigma\rangle_0
\hskip 0.75cm
=
\Big(\frac{1}{2} - 2\alpha\Big)\,
|\sigma\sigma\rangle_0
\\
&&
{\cal H}(\alpha)
|\rho\rho\rangle_0
\hskip 0.83cm
=
-\frac{3}{2}\,
|\rho\rho\rangle_0\ .
\end{eqnarray}

Since the energy eigenvectors of this generalized problem
are exactly the basis states we used for the
tetrahedron, the neutron scattering structure factors for the
S$_{\, tot} = 0$ to S$_{\, tot} = 1$ transitions are unchanged.
In this system however all these levels are nondegenerate, so
unlike the pure tetrahedron problem one encounters
no structure factor ambiguities due to an arbitrary choice
between degenerate basis states.

\subsubsection{Rectangular Tetramer}

The rectangular tetramer, shown in
Fig.\ref{123tetramerdia}b, has (12) and (34) dimers of
interaction strength J coupled by interactions
of strength $\alpha$J between ion pairs (13) and (24).
The Hamiltonian is
\be
{\cal H} =
{\J }
\Big(
\vec {\rm S}_{1}\cdot\vec {\rm S}_{2} +
\vec {\rm S}_{3}\cdot\vec {\rm S}_{4} +
\alpha
\big(
\vec {\rm S}_{1}\cdot\vec {\rm S}_{3} +
\vec {\rm S}_{2}\cdot\vec {\rm S}_{4}
\big)
\Big).
\label{Hrectangle}
\ee
This Hamiltonian is already diagonal on the S$_{\, tot} = 1,2$
dimer-pair basis states of
Eqs.(\ref{Seq2tetramerbasis},\ref{Seq1tetramerbasis}); the
energy eigenvalues are
\begin{eqnarray}
\E_{2\phantom{,3}} & = &+\frac{(1+\alpha)}{2}\, \J
\\
\E_{1,3} & = &+\frac{(1-\alpha)}{2} \, \J
\\
\E_{1,2} & = & -\frac{(1-\alpha)}{2}\, \J
\\
\E_{1,1} & = & -\frac{(1+\alpha)}{2}\, \J\ .
\label{RectangleESeq12}
\end{eqnarray}
The $2\times 2$ Hamiltonian matrix in the
S$_{tot} = 0$ subspace spanned by Eq.(\ref{Seq0tetramerbasis})
is
\be
{\cal H}_{\S_{tot}=0}  =
\frac{1}{2}\,
{\rm J}
\left[
\begin{array}{cc}
\ 1 - 2\alpha  \
& -\sqrt{3}\alpha
\\
-\sqrt{3}\alpha
&
\ -3 \
\end{array}
\right],
\label{RectSeq0Hmatrix}
\ee
which has the eigenvalues
\be
{\rm E}_{0, \{ {1\atop 2}\} } =
\Big(
-\frac{1}{2} - \frac{1}{2} \alpha  \mp
\sqrt{1 - \alpha + \alpha^2}
\, \Big) \, {\J }\ .
\label{ESeq0Rect}
\ee

The specific heat and susceptibility of the rectangular tetramer
may be evaluated using these energy levels and the general formulas of
Eqs.(\ref{C},\ref{chi}). The susceptibility is given in Table~\ref{chi_table}.
Although the specific heat is straightforward to evaluate, the resulting
expression is too lengthly to tabulate here.

The neutron scattering structure factors of the rectangular tetramer are
especially interesting because the ground state is a linear combination
of the dimer-pair basis states; this mixing leads to coupling-dependent
structure factors. The S$_{tot} = 0$ ground state of the rectangular
tetramer is a linear superposition of unexcited and doubly-excited dimer
pairs,
\be
 |\Psi_{0,1}\rangle =
-\sin(\theta_0) |\sigma\sigma\rangle_0 +
\cos(\theta_0) |\rho\rho\rangle \ ,
\ee
where the mixing angle $\theta_0$ between these basis states satisfies
\be
\tan(\theta_0) = - \frac{\sqrt{3}\,\alpha/2}{1-\alpha/2 +
\sqrt{1-\alpha+\alpha^2}} \ .
\ee

The matrix elements of the neutron scattering transition operator $V_\alpha$
of Eq.(\ref{Va_defn}) between the ground state $|\Psi_{0,1}\rangle$
and the three S$_{\, tot} = 1$ final states $|\Psi_{1,1\dots 3}\rangle$
may then be determined using the results of
Eqs.(\ref{tetramer_me2}-\ref{tetramer_me9}). Specializing to
S$_{z\, tot} = +1$ final states for illustration, these matrix elements
are
\bd
{}_{1,1}\langle \Psi_{1,n} | V_+ | \Psi_{0,1}\rangle =
\hskip 4cm
\ed
\be
\begin{cases}
\frac{(C_0 + S_0/\sqrt{3})}{2\sqrt{2}}\ (f_1 - f_2 - f_3 + f_4),
&\text{n = 1}
\\
\frac{(C_0 + S_0/\sqrt{3})}{2\sqrt{2}}\ (f_1 - f_2 + f_3 - f_4),
&\text{n = 2}
\\
\hskip0.95cm -\frac{S_0}{\sqrt{6}}\ (f_1 + f_2 - f_3 - f_4),
&\text{n = 3} \ ,
\end{cases}
\ee
where $C_0$ and $S_0$ are $\cos(\theta_0)$ and $\sin(\theta_0)$
respectively.

\begin{figure}[t]
\includegraphics[width=3.4in]{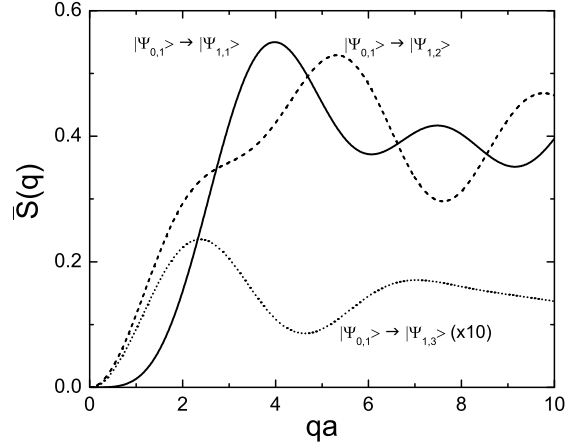}
\caption{Powder average, unpolarized structure factors
${\bar S(q)}$ for the excitation
of the three S$_{\, tot} = 1$ excited states of the rectangular
spin tetramer from the S$_{\, tot} = 0$ ground state $|\Psi_{0,1}\rangle$.
The interdimer coupling strength is $\alpha = 0.3$, and the side length
ratio is $b/a = 1.5$. The structure factor for the transition to
the third state is scaled up by a factor of 10 for visibility.}
\label{rectfig1}
\end{figure}

\begin{figure}[ht]
\includegraphics[width=3.4in]{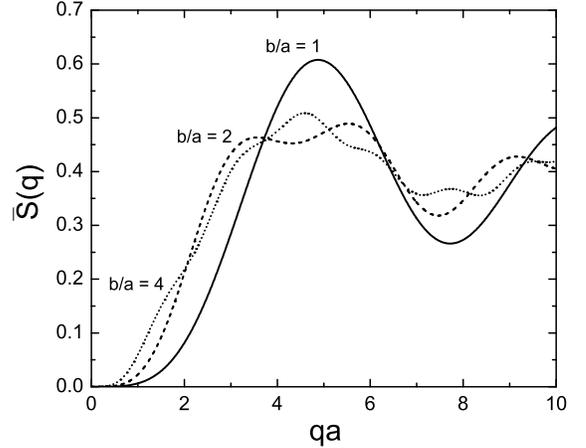}
\caption{Variation of the ${\bar S(q)}$ structure factors for excitation
of the lowest S$_{\, tot} = 1$ state of the rectangular
spin tetramer with side length ratio $b/a$.
This illustrates the use of powder structure factors in establishing
the internal geometry of spin clusters.
The dimer coupling is $\alpha = 0.3$,
however this only affects the overall normalization.}
\label{rectfig2}
\end{figure}

On converting these matrix elements to structure factors using
Eqs.(\ref{Str1},\ref{Strunpoldefn}) we find
different functional forms for each transition for both the single crystal
and powder average results, as a result of the different weight factors and
the three distinct ion pair separations.
The powder average, unpolarized structure factors are
\bd
{\bar S}^{\Psi_{0,1}\to\Psi_{1,n}}(q) =
\hskip 4cm
\ed
\be
\begin{cases}
\frac{(C_0 + S_0/\sqrt{3})^2}{2}
(1 - j_0(qa) - j_0(qb) + j_0(qc)),
&\text{n = 1}
\\
\frac{(C_0 + S_0/\sqrt{3})^2}{2}
(1 - j_0(qa) + j_0(qb) - j_0(qc)),
&\text{n = 2}
\\
\hskip 1.07cm \frac{2}{3}\,S_0^2\,
(1 + j_0(qa) - j_0(qb) - j_0(qc)),
&\text{n = 3}
\end{cases}
\ee
where $c = \sqrt{a^2+b^2}$.
In Fig.\ref{rectfig1} we show these structure factors for a case
with moderate interdimer coupling $(\alpha = 0.3)$ for a rectangle
of side ratio $b/a = 1.5$. The transition to the highest
S$_{\, tot} = 1$ excited state $|\Psi_{1,3}\rangle$ is much weaker than
the other two, and so is multiplied by a factor of 10 in the figure
for visibility.

Note that the excitation of the highest S$_{\, tot} = 1$ state
$|\Psi_{1,3}\rangle$, which is a doubly-excited dimer
($|\sigma\sigma\rangle$), is only possible because the ground state
has an $O(\alpha)$ excited component, in addition to the dominant
``bare'' $|\rho\rho\rangle$ basis state. The weakness of the
$|\Psi_{1,3}\rangle$ signal is because the structure factor is proportional
to the nonleading ground state amplitude squared, so that it is
$O(\alpha^2)$. The observation of similar ``non-valence state'' transitions
which are forbidden at $O(\alpha^0)$ should allow direct experimental tests
of the ``interaction'' terms in quantum spin Hamiltonians such as this one.
The strength alone is a sensitive measure of $\alpha$, and the spatial
modulation of $\bar S(q)$ is clearly different from the
$O(\alpha^0)$ transitions that dominate the structure factors of the
lower-lying S$_{\, tot} = 1$ states.

The detailed $q$-dependence of the structure factors can be used
as a ``fingerprint'' to test whether a given structure is indeed
the magnetically active system.
As an example, in Fig.\ref{rectfig2} we show that the detailed form of the
structure factor to the first S$_{\, tot} = 1$ state shows significant
variation with the ratio $b/a$.
(Values of $b/a = $ 1, 2 and 4 are shown.) This type of
dependence could be used to establish the geometry of a magnetic
subsystem, or to check powder neutron scattering results against the
geometry of a proposed spin system.

\subsubsection{Linear Tetramer}

The linear tetramer 
consists of two dimers with internal magnetic couplings of strength J,
with a single interdimer coupling of strength 
$\alpha$J between the two adjacent end spins (see Fig.\ref{123tetramerdia}). 
The term ``linear''
refers only to the pattern of magnetic couplings; the actual
spatial geometry of our linear tetramer is not assumed to be a straight line.
The ``Clemson tetramer'' NaCuAsO$_4$ \cite{Ulu03}
is a recent example of a possible
``linear tetramer'' that does not have a true collinear dimer geometry.

The linear tetramer Hamiltonian matrix is also relatively simple in
the dimer pair basis of
Eqs.(\ref{Seq2tetramerbasis}-\ref{Seq0tetramerbasis}).
The single S$_{tot}=2$ basis state
$|(\sigma\sigma)\rangle_2$ is diagonal, with energy
\be
\E_2 = \Big(\frac{1}{2} + \frac{1}{4} \alpha\Big)\, {\J } \ .
\ee
The S$_{tot}=1$ basis state $|(\rho\sigma)_A\rangle$ is also diagonal,
with energy
\be
\E_{1,\, 2} = -\Big(\frac{1}{2} - \frac{1}{4} \alpha\Big)\, {\J } \ .
\ee
The remaining S$_{tot}=1$ and S$_{tot}=0$
two-dimensional Hamiltonian matrices are
\be
{\cal H}_{\S_{tot}=1}  =
\frac{1}{2} \, {\rm J}
\left[
\begin{array}{cc}
\ 1 - \frac{1}{2} \alpha  \
& \alpha
\\
\alpha
&
\ -1  - \frac{1}{2} \alpha \
\end{array}
\right]
\label{LinSeq1Hmatrix}
\ee
and
\be
{\cal H}_{\S_{tot}=0}  =
\frac{1}{2}\, {\rm J}
\left[
\begin{array}{cc}
\ 1 - \alpha  \
& \frac{\sqrt{3}}{2}\alpha
\\
\frac{\sqrt{3}}{2}\alpha
&
\ - 3 \
\end{array}
\right]
\label{LinSeq0Hmatrix}
\ee
with eigenvalues
\be
{\rm E}_{1, \{ {1\atop 3}\} } =
-
\Big(
\frac{1}{4} \alpha  \pm
\frac{1}{2}\sqrt{1 + \alpha^2}\;
\Big)\, {\rm J}
\label{ESeq1Lin}
\ee
and
\be
{\rm E}_{0, \{ {1\atop 2}\} } =
-\Bigg(
\frac{1}{2} + \frac{1}{4} \alpha  \pm
\sqrt{1 - \frac{1}{2} \alpha + \frac{1}{4} \alpha^2}
\; \Bigg) \; {\rm J}
\label{ESeq0Lin}
\ee
respectively.

The susceptibility of the linear tetramer, which follows from
these energy levels and Eq.(\ref{chi}), is given in
Table~\ref{chi_table}. As with the rectangular tetramer,
the expression we find for the 
specific heat is too lengthly to tabulate here.

The neutron scattering structure factors from the linear tetramer
S$_{\, tot} = 0$
ground state $|\Psi_{0,1}\rangle$ to the three
S$_{\, tot} = 1$ excited states $|\Psi_{1,1\dots 3}\rangle$
may be calculated
using the same techniques we applied to the rectangular tetramer.
The results are somewhat more complicated, since two of the
S$_{\, tot} = 1$ linear tetramer states are rotated between the
dimer-pair basis states, in addition to the ground state basis
rotation we found for the rectangular tetramer.
The energy eigenvectors in these sectors
are the superpositions of dimer-pair basis states
given in Table~\ref{wavefunctions2}, with mixing angles
that satisfy
\begin{eqnarray}
&&
\tan(\theta_1) = \frac{\alpha}{1+\sqrt{1+\alpha^2}} \ ,
\label{lintettheta1}
\\
&&
\tan(\theta_0) =
\frac{\sqrt{3}\,\alpha/4}{1 - \alpha/4 +\sqrt{1 - \alpha/2 +\alpha^2/4}} \ .
\label{lintettheta0}
\end{eqnarray}
This more complicated
basis mixing pattern introduces a new feature, which is that the
functional forms of the structure factors for the two mixed
S$_{\, tot} = 1$ states, $|\Psi_{1,1}\rangle$ and
$|\Psi_{1,3}\rangle$, depend
on the dimer coupling $\alpha$. (In the rectangular tetramer
system discussed previously we found that changing the dimer coupling
$\alpha$ only changed the overall normalization of the structure factors,
not their detailed $q$ dependence.)

We will give explicit results for the first transition,
$|\Psi_{0,1}\rangle\to|\Psi_{1,1}\rangle$,
and then simply quote the results for the
two remaining final states.
The matrix element of the neutron scattering transition operator
$V_a$ is
\be
{}_{11} \langle \Psi_{1,1} | V_+ |\Psi_{0,1}\rangle
=
c_{14} \, (f_1 - f_4)
+
c_{23} \, (f_2 - f_3)
\ee
where
\be
\left\{
\begin{array}{l}
c_{14} \\
c_{23} \\
\end{array}
\right\}
=
\pm
\frac{(C_0 + S_0/\sqrt{3})\, C_1}{2\sqrt{2}}
+
\frac{S_0 S_1 }{\sqrt{6}}
\ee
and $C_{0,1}$ and $S_{0,1}$ are respectively
cos and sin of the linear tetramer
basis state mixing angles, which were defined in
Eqs.(\ref{lintettheta1},\ref{lintettheta0}).

\begin{figure}[ht]
\includegraphics[width=3.4in]{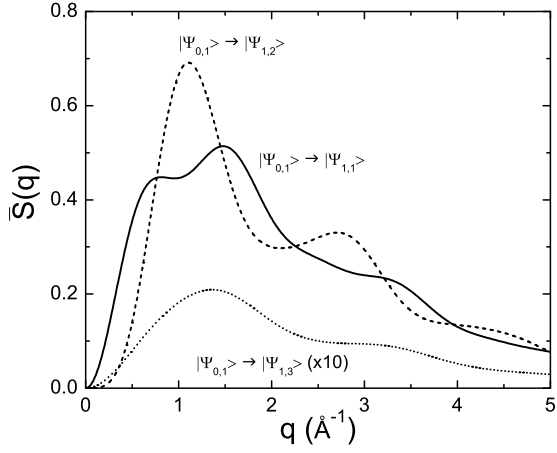}
\caption{Powder average, unpolarized structure factors
${\bar S(q)}$ for the excitation
of the three S$_{\, tot} = 1$ states of the linear tetramer,
including the ionic form factor,
with magnetic coupling ratio $\alpha = 0.4$ and ion positions
taken from the NaCuAsO$_4$ structure \cite{Ulu03}. The structure factor
for the transition to the third state is scaled up by
a factor of 10 for visibility.}
\label{clems_fig}
\end{figure}

\eject
The resulting unpolarized powder structure factor for this transition,
using Eqs.(\ref{Strunpoldefn},\ref{Strpowavg}), is
\bd
{\bar S}^{\Psi_{0,1}\to \Psi_{1,1}} =
\hskip 4cm
\ed
\bd
2
\Big(
c_{14}^{\; 2}
\big( 1 - j_0(q r_{14})\big)
+
c_{23}^{\; 2}
\big( 1 - j_0(q r_{23})\big)
\ed
\be
\hskip 1cm
+ 2\, c_{14} \, c_{23}\,
(j_0(q r_{12}) -  j_0(q r_{13}))
\Big) .
\label{linstr1}
\ee

The transition from the ground state to the second linear tetramer
S$_{\, tot} = 1$ excited state is given by a matrix element we encountered
previously in the rectangular tetramer problem, except for a change in
spatial geometry.
The result for the unpolarized powder structure factor with completely
general ion positions is
\bd
{\bar S}^{\Psi_{0,1}\to\Psi_{1,2}}(q) =
\hskip 4cm
\ed
\bd
\frac{(C_0 + S_0/\sqrt{3})^2}{2}
\Big(
1 + \frac{1}{2}
\big(
  - j_0(qr_{12}) - j_0(qr_{13}) + j_0(qr_{14})
\ed
\be
  + j_0(qr_{23}) - j_0(qr_{24}) - j_0(qr_{34})
\big)
\Big).
\label{linstr2}
\ee

\begin{figure}[h]
\vskip 0.8cm
\includegraphics[width=3.2in]{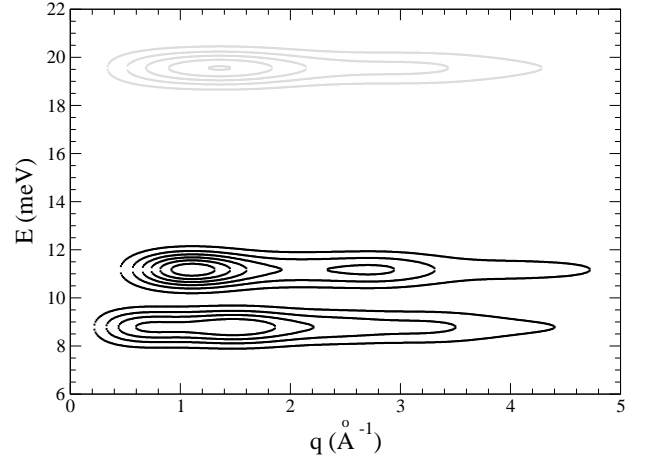}
\caption{Contours of equal intensity of the powder average,
unpolarized structure factors of the linear tetramer,
with parameters appropriate to NaCuAsO$_4$,
as in Fig.\ref{clems_fig}. The contours are normalized
to the peak of the first transition, and intervals of 0.2
in intensity are shown for the
two lower states. The much weaker third transition is displayed with
intervals of 0.01.}
\label{clemscontour}
\end{figure}

The structure factor for the third S$_{\, tot} = 1$ state
$\Psi_{1,3}$ may be found from the $\Psi_{1,1}$ structure factor of
Eq.(\ref{linstr1}) with the simple substitutions
$(C_1\to S_1), (S_1\to - C_1)$.

We will illustrate the predicted structure factors for the linear tetramer
assuming parameters appropriate for the candidate material
NaCuAsO$_4$ \cite{Ulu03,Nag03}. The ion pair separations are
$r_{12} = r_{34} = 3.641$~\AA,
$r_{13} = r_{24} = 3.863$~\AA\,
$r_{14} =          6.814$~\AA\ 
and
$r_{23} = 3.151$~\AA.
The copper positions are consistent with planarity, although
our results do not require this assumption.
There are indications
of three S$_{\, tot} = 1$ levels in this material
from a recent inelastic neutron scattering experiment, with
energies of approximately 9, 11 and 18~meV \cite{Nag03};
in the linear tetramer model this
suggests parameters of J $\approx 10$~meV
and $\alpha \approx 0.4$, which we will assume here.
(The structure factors only depend on $\alpha$.)
We also incorporated a simple Cu$^{2+}$ ionic form factor,
$F(q) = 1/(1+q^2/q_0^2)^3$ with $q_0 = 8.0$~\AA$^{-1}$,
which agrees with the online ILL
Cu$^{2+}$ form factor \cite{ILLsite} to $\lesssim 0.5$~\% over the
range of $q$ shown.
Our results are shown in Fig.\ref{clems_fig}; characteristic features
include the displaced relative maxima of the intensities of the
two lower states, and the much weaker transition to the highest
S$_{\, tot} = 1$ state. The results for the two lower states are
rather insensitive to $\alpha$.
The overall scale of the $|\Psi_{1,3}\rangle$ structure factor
however is quite sensitive to $\alpha$, and scales approximately
as $\alpha^2$. A measurement of the
relative strength of these transitions would provide a useful
determination if $\alpha$, which could be compared with the
value extracted from the energy levels. (In principle the
susceptibility could also be used to determine $\alpha$,
but we have found that it has a
rather weak $\alpha$ dependence in this system.)

In Fig.\ref{clemscontour} we show these results in a contour plot,
approximately as would be observed in a neutron scattering experiment.
(Our results should be multiplied by the energy-dependent factor
$k'/k$ of Eq.(\ref{cross_sec}) for a direct comparison with experiment.)
To generate this plot we have
convolved a Gaussian energy resolution function,
$\exp(-(E-E_i)^2/2\sigma_E^2)$ with $\sigma_E = 0.5$~meV,
with the structure factors to the three S$_{tot}=1$ states.
The intensities are shown relative to the maximum excitation intensity of
the transition to the lowest S$_{tot}=1$ state,
$|\Psi_{0,1}\rangle \to |\Psi_{1,1}\rangle$. Note the characteristic
strong peak in intensity of the second transition,
$|\Psi_{0,1}\rangle \to |\Psi_{1,2}\rangle$, near $1.1$~\AA$^{-1}$.
Comparison of these general features with the data of Nagler {\it et al.}
\cite{Nag03}
suggests that the linear tetramer model does indeed give a realistic
description of neutron scattering from NaCuAsO$_4$.

\section{Future Applications}

In the previous section we presented results for bulk
thermodynamic and magnetic properties
of various dimer, trimer and tetramer molecular magnets with
S~=~1/2 ions. We also derived the inelastic neutron scattering
structure factors for these systems; inelastic neutron scattering
is very useful as a local probe of magnetic interactions at the
atomic scale. These results clearly have many possible applications
to real materials. In this section we discuss some examples
of materials that are thought to be realizations of S~=~1/2 dimer,
trimer and tetramer molecular magnets, and describe how our results
could be useful in future experimental investigations.
We also discuss possible extensions of this work, which will
be useful in interpreting experimental data on these and related
magnetic materials under more general circumstances.

The S~=~1/2 spin dimer is the simplest of all spin clusters.
It provides a textbook case for studies of finite spin systems more generally,
since physical observables for the dimer can often be derived as closed form
analytic expressions. This is the case for the specific heat, susceptibility
and neutron scattering structure factors presented here.
Vanadyl hydrogen phosphate, VO(HPO$_4$)$\cdot$0.5H$_2$O, is a well known
example of an S = 1/2 spin dimer material 
\cite{Joh84,Ten97,Koo04}, 
and
some of the results tabulated here have already been used in interpreting data
on this material. In particular, the magnetic susceptibility was originally
used to determine the exchange constant J, and inelastic neutron scattering
was used to test the simple dimer model and establish which pair of
V$^{4+}$ ions forms the dimer \cite{Ten97}. This experiment was a dramatic
success for inelastic neutron scattering, as the previously assumed
V-V dimer pair was shown to have been misidentified.

Many examples of S=1/2 ion trimers have been reported in the literature.
These systems are interesting in that the ground state is
(ideally) degenerate, and must exhibit ferromagnetism.
(S$_{tot} > 0$ for any energy eigenstate of
an isotropic magnetic Hamiltonian with half-integer ion spins and
an odd number of ions.)
Heisenberg trimers with antiferromagnet pair interactions
are also of interest because they are the
simplest isotropic spin systems which experience frustration.
One example of an S=1/2 trimer is
Cu$_3$(O$_2$C$_{16}$H$_{23}$)$_6\cdot 1.2$C$_6$H$_{12}$
\cite{Cag03a,Cag03b},
which has an equilateral Cu$^{2+}$ triangle
with a Cu$^{2+}$ - Cu$^{2+}$ separation
of 3.131 \AA.
Recent Electron Paramagnetic Resonance (EPR) measurements
show that the ground state of this material consists of a
twofold-degenerate S$_{tot} = 1/2$ level; this is in accord with
expectations for a general isotropic trimer antiferromagnet with
S = 1/2 ions, but not with the perfect equilateral (symmetric) case,
in which the ground state is a quartet of two degenerate
S$_{tot} = 1/2$ levels.
The gap to the S$_{tot} = 3/2$ excited level is estimated
from both EPR and susceptibility data to be
28~meV~\cite{Cag03a,Cag03b}.
There are indications from the EPR studies that
this fourfold degeneracy has been lifted by additional, nonisotropic
interactions~\cite{Cag03b}.
Investigation of the level structure and structure factors
in this apparently symmetric trimer material would be a very interesting
exercise for a future inelastic neutron scattering experiment, especially
if large single crystals are available.

Another recent example of an S=1/2 trimer is the ``Na$_{9}$-2" material
of Kortz {\it et al.} \cite{Kor04},
Na$_9$[Cu$_3$Na$_3$(H$_2$O)$_9$ 
($\alpha$-AsW$_9$O$_{33}$)$_2$]$\cdot$26H$_2$O,
which
contains an equilateral Cu$^{2+}_3$ trimer with a susceptibility consistent
with equal Heisenberg interactions of J~$\approx 0.35$~meV.
Inelastic neutron scattering from a powder sample
of this material should show the S$_{tot}=3/2$
excited level, with a structure factor proportional to $(1-j_0(qa))$.
With a single crystal sample it might be possible to separate the
transitions from the two (nearly?) degenerate S$_{tot}=1/2$ levels
to the S$_{tot}=3/2$ excited level.

The two ``V$_{6}$" materials
of Luban~{\it et al.}~\cite{Lub02},
(CN$_3$H$_6$)$_4$
Na$_2$[H$_4$V$_6$O$_8$(PO$_4$)$_4$
((OCH$_2$)$_3$ CCH$_2$OH)$_2$]
$\cdot$14H$_2$O
and
Na$_6$[H$_4$V$_6$O$_8$(PO$_4$)$_4$
((OCH$_2$)$_3$CCH$_2$OH)$_2$]$
\cdot$18H$_2$O,
contain pairs of
(presumably weakly coupled) V$_{3}$ spin trimers that are
respectively isosceles and general triangular systems. The isosceles V$_{6}$
material was used as an example of single-crystal inelastic neutron scattering
structure factors in this paper.

The ``V$_{15}$" material 
K$_6$[V$_{15}$As$_6$O$_{42}$(H$_2$O)]$\cdot$8H$_2$O
is an
example of an S=1/2 trimer in a more complicated magnetic geometry.
This material has a
frustrated V$_3$ triangle sandwiched between two nonplanar antiferromagnetic
V$_6$ hexagons.
The low-temperature magnetic properties are
dominated by the V$_3$ triangle; other magnetic interactions become important
at elevated temperatures
\cite{Bar92,Mue88,Gat91,Cha02,Cha04,Cho03}.
In addition to distinguishing between direct vanadium-vanadium and superexchange
pathways involving the upper and lower hexagons, neutron scattering
was used to probe the magnetic structure, finding two nearly degenerate 
S$_{tot}$ = 1/2 ground states (with 0.035~meV splitting) 
and an S$_{tot}$ = 3/2 excited state \cite{Cha02}. The availability of
large single crystals of this material suggests that it might be an interesting
candidate for inelastic neutron scattering studies. 

Clearly, the analytical expressions we have presented here
for spin-trimer thermodynamic properties and inelastic neutron
scattering amplitudes have wide potential application,
and should be useful in particular for interpreting
the results of future inelastic neutron scattering experiments
on spin-trimer molecular magnets.

Examples of tetramer systems with S = 1/2 ions include sodium copper
arsenate, NaCuAsO$_4$ \cite{Ulu03,Nag03}, which we used as an
illustration of the evaluation of inelastic neutron
scattering structure factors in the previous section. The neutron
scattering data of Nagler {\it et al.} \cite{Nag03}  supports a
model of this material as an open-chain tetramer, with
antiferromagnetic Heisenberg bonds of alternating strength.
Transitions from the S$_{tot} = 0$ ground state to all three
S$_{tot} = 1$ triplet excited states have been observed on a powder
sample \cite{Nag03}, 
and the relative intensities and the $q$-dependence of the 
powder average structure factors appear to be approximately consistent
with our predictions for the open-chain model. As we have given detailed analytic
predictions for the neutron structure factor for these transitions,
a comparison with data from a high-statistics experiment on a larger
powder sample should be straightforward. 

Additional examples of 
S~=~1/2 tetramers are found in vanadium materials containing the
[V$_{12}$As$_8$O$_{40}$(H$_2$O)]$^{4-}$ cluster anion. Basler {\it
et al.} \cite{Bas02} have recently reported studies of the magnetic
properties of three such materials,
Na$_4$[V$_{12}$As$_8$O$_{40}$(H$_2$O)]$\cdot$23H$_2$O,
Na$_4$[V$_{12}$As$_8$O$_{40}$(D$_2$O)] 
$\cdot$16.5D$_2$O 
and 
(NHEt$_3$)[V$_{12}$As$_8$O$_{40}$(H$_2$O)]$\cdot$H$_2$O. 
These materials have three stacked V$_{4}$
tetramers, but are mixed-valent (V$^{4+}_8$V$^{+5}_4$). The middle
tetramer dominates the magnetic properties. This tetramer is
anti- ferromagnetic and close to square, with exchange constants of
$\approx 1.5$~meV (inferred from energy levels established by EPR
and inelastic neutron scattering). The Basler {\it et al.} study is
a very nice illustration of the combined use of bulk magnetic
properties and inelastic neutron scattering to characterize magnetic
materials, as we advocate in this work. Additional studies of this
already well characterized material might involve an inelastic
neutron scattering study of a single crystal, which could be used to
test the detailed orientation dependence expected for the structure
factor for each of the observed magnetic transitions to excited
states, given their fitted magnetic Hamiltonian. Since this
Hamiltonian includes anisotropies, a study using polarized neutrons
could provide additional useful information.

There are several interesting questions which were not considered
in detail in this paper
that would be appropriate for future research on finite spin clusters.
Consideration of higher ionic
spin is one obvious generalization of this work.
Several examples of uncompensated molecular magnets (which have ground states
with nonzero spin) may be found in relatively simple
higher-spin materials.
One example is the first cobalt molecular
magnet \cite{Yang}, Co$_4$(NC$_5$H$_4$H$_2$CO)$_4$(CH$_3$OH)$_4$Cl$_4$.
This material consists of four S = 3/2 Co$^{2+}$ ions and four ligand-related
oxygen atoms situated on the corners of a cube, with a ferromagnetic
S$_{tot} = 6$ ground state.
The magnetic exchange constants have been estimated from fits to
magnetization curves, and are at the meV scale \cite{Yang}.
The magnetic Co$^{2+}$ ions in this material form a tetramer
with important tetrahedral and dimer magnetic interactions \cite{Yang},
which would be an important case for future neutron scattering
studies, especially with large single crystals.
Another example of a molecular magnet with higher ion
spin is the chromium magnet
[Cr$_4$S(O$_2$CCH$_3$)$_8$(H$_2$O)$_4$](NO$_3$)$_2$$^.$H$_2$O,
which has four ferromagnetically coupled S~=~3/2 Cr$^{3+}$ ions
arranged in a nearly regular tetrahedron, with an
S$_{tot} = 6$ ground state. Furukawa {\it et al.}
\cite{Fur00}  determined the exchange constant for this
material from the susceptibility, and predict a gap to the first
S$_{tot} = 5$ excited state of ca.~15~meV. Observation of this
excitation using inelastic neutron scattering should be a
straightforward exercise, and the intensities should agree well
with theoretical expectations of the Heisenberg model, since this model
gives a reasonably good description of the bulk magnetic properties.

Extension of this work to mixed-valent spin clusters would also be interesting,
since many examples of these are known, including systems with
magnetized ground states.
One example is the ``Mn$_4$" material
[Mn$_4$O$_3$Cl$_4$(O$_2$CEt)$_3$(py)$_3$]$_2\cdot 2$C$_6$H$_{14}$
studied by
Hill $et~al.$ \cite{Hil03}, which has a mixed-valent spin tetramer
consisting of a triangle of S~=~2 Mn$^{3+}$ ions with an apical
S~=~3/2 Mn$^{4+}$, and a ground-state spin of
S$_{tot}=9/2$. Although the interest in this material as a
molecular magnet is largely due to the weak coupling between pairs
of Mn$_4$ clusters, the magnetic Hamiltonian within a single
Mn$_4$ cluster could be tested by powder inelastic neutron scattering.

Another interesting theme for future studies is
the effect of finite temperatures on inelastic neutron scattering;
although increasing temperatures are usually associated with weaker
inelastic transitions, the finite Hilbert space of a spin cluster
implies that magnetic transitions will weaken in a simple, known manner
according to their Boltzman factors, and will approach finite limits
at high temperatures (provided that the magnetic Hamiltonian remains
valid). 
Since moderate temperatures (on the scale of the magnetic excitations)
will significantly populate excited levels,
it may also be possible to observe inelastic transitions from excited levels
that are inaccessible at low temperatures.

Another important topic which we briefly alluded to in the text is the issue of
magnetic interactions between spin clusters; these interactions will broaden
the discrete levels assumed here into bands, which will be observable if the
intercluster interactions are sufficiently large.

Finally, the generalization of our results to non-Heisenberg interactions,
and the determination of these interaction parameters through
polarized inelastic neutron scattering experiments, would be an
especially interesting and important extension of the work
presented here.

\section{Summary and Conclusions}

In this paper we have evaluated several thermodynamic and neutron scattering
observables that characterize the magnetic behavior of
finite quantum spin systems.
After an introduction that gives results applicable to
the general case, we specialized to clusters
of S=1/2 ions with a Heisenberg interaction between ion pairs.
We considered dimer, trimer and tetramer systems with various
magnetic interaction strengths, and evaluated the
magnetic specific heat, the susceptibility,
and the inelastic neutron scattering structure factor for these systems.
The structure factor was derived both for single crystals and
for the powder average case.
Our results for the neutron scattering structure factor
show that accurate intensity measurements of inelastic neutron scattering
cross sections from a powder can be useful in establishing the
spatial geometry of an assumed set of interacting magnetic ions.
The linear spin-tetramer candidate NaCuAsO$_4$ was considered as an example,
and we found that the observed inelastic powder pattern for excitation of the
two lowest S$_{tot}=1$ excited levels 
is indeed consistent with the predictions of
the linear tetramer model.
We also considered inelastic neutron scattering from single crystals,
and found dramatic angular dependence that could be used in future
experiments as sensitive tests of the assumed magnetic Hamiltonian.
We concluded with a discussion of specific materials that might
be studied using our results, and suggested future extensions of our
work to more general systems.

\section{Acknowledgements}

This project was supported by the Petroleum Research Fund
administered by the American Chemical Society (PRF-AC 38164)
and the Joint Institute for Neutron Sciences. TB also acknowledges 
support from the Neutron Sciences Consortium of the 
University of Tennessee. We are grateful to C.C.Torardi
for providing a sample of the spin dimer
VO(HPO$_4$)$\cdot$0.5H$_2$O,
J.R.Thompson for measuring the susceptibility,
and S.E.Nagler for useful communications and access to unpublished data 
on NaCuAsO$_4$.

\renewcommand{\theequation}{A.\arabic{equation}}
\setcounter{equation}{0}  
\section*{APPENDIX: Tetramer basis states and matrix elements}  

Since there is a natural separation of the rectangle and linear
tetramer systems into dimer components, it is useful to 
introduce a
$|(12)(34)\rangle$ dimer basis
to represent tetramer energy eigenvectors. The dimer basis
states are
\be
|\rho\rangle = \frac{1}{\sqrt{2}}\,
\Big(|\!\upa\dna\,\rangle - |\!\dna\upa\,\rangle\Big)
\ee
and
\begin{eqnarray}
|\sigma(m)\rangle =
\left[
\begin{array}{cl}
|\!\upa\upa\,\rangle
&
m = +1
\\
\\
\frac{1}{\sqrt{2}}\, \Big(|\!\upa\dna\,\rangle + |\!\dna\upa\,\rangle\Big)
&
m = 0
\\
\\
|\!\dna\dna\,\rangle
&
m = -1
\\
\end{array}
\right]_{\ . }
\label{tetramerbasis}
\end{eqnarray}
These are combined as Clebsch-Gordon series to form tetramer basis states of definite
total spin and symmetry, which are
$|(\sigma\sigma)_S\rangle,\,   \S_{tot} = 0,2$;
$|(\sigma\sigma)_A\rangle,\,   \S_{tot} = 1$;
$|(\rho\sigma)_{S,A}\rangle,\, \S_{tot} = 1$;
$|\rho\rho\rangle, \, \S_{tot} = 0$.
In the interest of clarity we will occasionally specify the total
spin of one of these basis state with a subscript; thus
$|\sigma\sigma\rangle_0$ refers to the
$|(\sigma\sigma)_S\rangle$ state with
$\S_{tot} = 0$.

Using these states as basis vectors reduces the
16-dimensional full tetramer Hilbert space
to 1-, 2- and 3-dimensional subspaces,
which are spanned by the basis sets
\be
\!
|\; \S_{tot} = 2  \rangle
\ \ \ \ = \ \
|\sigma\sigma\rangle_2
\\
\label{Seq2tetramerbasis}
\ee
\be
\qquad \!
\{\;
| \; \S_{tot} = 1  \rangle \;\} =
\left[
\begin{array}{c}
|\sigma\sigma\rangle_1
\\
\\
|(\rho\sigma)_S\rangle
\\
\\
|(\rho\sigma)_A\rangle
\\
\end{array}
\right]
\label{Seq1tetramerbasis}
\ee
\be
\quad \!
\{\; |\; \S_{tot} = 0  \rangle \;\} \ =
\left[
\begin{array}{c}
|\sigma\sigma\rangle_0
\\
\\
|\rho\rho\rangle
\\
\end{array}
\right]_{\ .}
\label{Seq0tetramerbasis}
\ee
Thus symmetry arguments alone determine the eigenvectors for one
level, and the eigenvectors for the remaining levels involve
at most $2\times 2$ and $3\times 3$ diagonalizations.
As we shall see, for the three tetramer models we consider here we
actually encounter at most $2\times 2$ diagonalization problems
using this basis.

These basis states are also convenient for determining
neutron scattering structure factors,
since they have relatively simple matrix elements of the
spin transition operator $V_a$ of Eq.(\ref{Va_defn}).
The complete set of matrix elements of $V_a$
(spherical components)
between single (12)-dimer basis states,
with $f_i = e^{i\vec k \cdot \vec x_i}$, is

\begin{eqnarray}
&&
\langle \rho\, | \, V_a \, | \rho \rangle    =  0
\\
&&
\langle \sigma(m)\, | \, V_a \, | \rho \rangle   =
\delta_{a,m}
\frac{f_1 - f_2}{2}
\\
&&
\langle \rho\, | \, V_a \, | \sigma(m) \rangle   =
-\delta_{a,-m}
\frac{f_1 - f_2}{2}
\\
&&
\langle \sigma(m') | \, V_a \, | \sigma(m) \rangle   =
-\delta_{m',m+a}
 \frac{f_1 + f_2}{2}\ .
\end{eqnarray}

These dimer results may be combined to give the complete set of
matrix elements of $V_a$ between tetramer basis states, which is all
that we require to determine all neutron scattering
structure factors for all the spin tetramer problems we consider.
These tetramer matrix elements (with explicit
S$_{\, tot}$ or S$_{\, tot}$, S$_{z\, tot}$ subscripts on the states
where required for clarity) are
\begin{eqnarray}
&&
\phantom{{}_{1,m}}
\langle \rho\rho\, | \, V_a \, | \rho\rho \rangle   =  0
\hskip 5cm
\label{tetramer_me1}
\\
&&
\nonumber
\\
&&
{}_{1,m}\langle (\rho\sigma)_S\, | \, V_a \, | \rho\rho \rangle   =
\nonumber
\\
&&
\hskip 3cm
\delta_{a,m}
\frac{f_1 - f_2 + f_3 - f_4}{2\sqrt{2}}
\label{tetramer_me2}
\\
&&
\nonumber
\\
&&
{}_{1,m}\langle (\rho\sigma)_A\, | \, V_a \, | \rho\rho \rangle   =
\nonumber
\\
&&
\hskip 3cm
\delta_{a,m}
\frac{f_1 - f_2 - f_3 + f_4}{2\sqrt{2}}
\label{tetramer_me3}
\\
&&
\phantom{{}_{1,m}}
\langle \sigma\sigma \, | \, V_a \, | \rho\rho \rangle   =  0
\hskip 5cm
\label{tetramer_me4}
\\
&&
\nonumber
\\
&&
\phantom{{}_{1,m}}
\langle \rho\rho \, | \, V_a \, | \sigma\sigma \rangle   =  0
\hskip 5cm
\label{tetramer_me5}
\\
&&
\nonumber
\\
&&
{}_{1,m}\langle (\rho\sigma)_S\, | \, V_a \, | \sigma\sigma \rangle_0   =
\nonumber
\\
&&
\hskip 3cm
- \delta_{a,m}
\frac{f_1 - f_2 + f_3 - f_4}{2\sqrt{6}}
\label{tetramer_me6}
\\
&&
\nonumber
\\
&&
{}_{1,m}\langle (\rho\sigma)_A\, | \, V_a \, | \sigma\sigma \rangle_0   =
\nonumber
\\
&&
\hskip 3cm
- \delta_{a,m}
\frac{f_1 - f_2 - f_3 + f_4}{2\sqrt{6}}
\label{tetramer_me7}
\\
&&
\nonumber
\end{eqnarray}

\begin{eqnarray}
&&
{}_{\phantom{,m}0}\langle \sigma\sigma\, | \, V_a \, | \sigma\sigma \rangle_0   = 0
\label{tetramer_me8}
\\
&&
\nonumber
\\
&&
{}_{1,m}\langle \sigma\sigma\, | \, V_a \, | \sigma\sigma \rangle_0   =
\nonumber
\\
&&
\hskip 3cm
\delta_{a,m}
\frac{f_1 + f_2 - f_3 - f_4}{\sqrt{6}}
\label{tetramer_me9}
\\
&&
\nonumber
\\
&&
{}_{\phantom{,m}2}\langle \sigma\sigma\, | \, V_a \,
| \sigma\sigma \rangle_0   = 0
\label{tetramer_me10}
\\
&&
\nonumber
\end{eqnarray}

The remaining matrix elements between pairs of S$_{tot}$~$=$~$1$ states 
and between S$_{tot}=1$ and S$_{tot}=2$ states, which were not required in this 
paper, may be evaluated similarly.

\vfill\eject

\begin{table*}
\caption{Energy eigenvectors and eigenvalues.}
\vskip 0.5cm
\begin{ruledtabular}
\begin{tabular}{lll}
Spin System
& Eigenvector $|\Psi_{\S_{tot}}\rangle$ (S$_{z\, tot} =$ S$_{\, tot}$)
& Energy
\\
\colrule \hline
&
\\
Dimer
&
$|\Psi_1\rangle = |\sigma(+1)\rangle =
|\!\upa\upa\,\rangle$
&
E$_1 = \phantom{-}\frac{1}{4}\,{\rm J}$
\\
\\
&
$|\Psi_0\rangle = |\rho\rangle =
\quad
\frac{1}{\sqrt{2}}
(|\!\upa\dna\,\rangle -
 |\!\dna\upa\,\rangle)$
&
E$_0 = -\frac{3}{4}\,{\rm J}$
\\
\\
\colrule \hline
\\
Symmetric Trimer
&
$|\Psi_{\frac{3}{2}}\rangle \ = \;
|\sigma(+3/2)\rangle = |\!\upa\upa\upa\,\rangle$
&
E$_{\frac{3}{2}} = \phantom{-}\frac{3}{4}\,{\rm J}$
\\
\\
&
$|\Psi_{\frac{1}{2},2}\rangle =
|\lambda(+1/2)\rangle =
\frac{1}{\sqrt{6}}
(   |\!\upa\dna\upa\,\rangle
+   |\!\dna\upa\upa\,\rangle
- 2 |\!\upa\upa\dna\,\rangle ) $
&
E$_{\frac{1}{2}} = -\frac{3}{4}\,{\rm J}$
\\
\\
&
$|\Psi_{\frac{1}{2},1}\rangle =
|\rho(+1/2)\rangle =
\frac{1}{\sqrt{2}}
(|\!\upa\dna\upa\,\rangle -
 |\!\dna\upa\upa\,\rangle)$
&
\\
\\
\colrule
\\
Isosceles Trimer
&
$|\Psi_{\frac{3}{2}}\rangle \ = \;
|\sigma(+3/2)\rangle $
&
E$_{\frac{3}{2}} = (\frac{1}{4} + \frac{1}{2}\alpha)\,{\rm J}$
\\
\\
&
$|\Psi_{\frac{1}{2},2}\rangle =
|\lambda(+1/2)\rangle $
&
E$_{\frac{1}{2},2} = (\frac{1}{4} - \alpha )\,{\rm J}$
\\
\\
&
$|\Psi_{\frac{1}{2},1}\rangle =
|\rho(+1/2)\rangle $
&
E$_{\frac{1}{2},1} = -\frac{3}{4}\,{\rm J}$
\\
\\
\colrule
\\
General Trimer
&
$|\Psi_{\frac{3}{2}}\rangle \ = \;
|\sigma(+3/2)\rangle $
&
E$_{\frac{3}{2}} \  = \frac{1}{4} ( 1 + \alpha_s )\,{\rm J}$
\\
\\
&
$|\Psi_{\frac{1}{2},2}\rangle =
+ \cos(\theta) |\lambda(+1/2)\rangle
+ \sin(\theta) |\rho(+1/2)\rangle
$
&
E$_{\frac{1}{2},2} =
\frac{1}{4}(
+
\sqrt{(2-\alpha_s)^2 + 3\, \alpha_d^2 }\,
- 1  - \alpha_s )\,
{\rm J}
$
\\
\\
&
$|\Psi_{\frac{1}{2},1}\rangle = -
\sin(\theta) |\lambda(+1/2)\rangle
+
\cos(\theta) |\rho(+1/2)\rangle
$
&
E$_{\frac{1}{2},1} =
\frac{1}{4}(
-
\sqrt{(2-\alpha_s)^2 + 3\, \alpha_d^2 }\,
- 1  - \alpha_s )\,
{\rm J}
$
\\
\\
\colrule \hline
\\
Tetrahedron
&
$|\Psi_2\rangle =
|\sigma\sigma\rangle_2 \ = |\!\upa\upa\upa\upa\,\rangle$
&
E$_2 =
\phantom{-}\frac{3}{2}\,{\rm J}$
\\
\\
&
$|\Psi_{1,3}\rangle = |\sigma\sigma\rangle_1 $
&
E$_{1} = -\frac{1}{2}\,{\rm J}$
\\
\\
&
$|\Psi_{1,2}\rangle = |(\rho\sigma)_S\rangle  $
&
\\
\\
&
$|\Psi_{1,1}\rangle = |(\rho\sigma)_A\rangle  $
&
\\
\\
&
$|\Psi_{0,2}\rangle = |\sigma\sigma\rangle_0  $
&
E$_0 = -\frac{3}{2}\,{\rm J}$
\\
\\
&
$|\Psi_{0,1}\rangle = |\rho\rho\rangle  $
&
\\
\\
\colrule
\\
Rectangular tetramer
&
$|\Psi_2\rangle \ = \ |\sigma\sigma\rangle_2 $
&
E$_2 = (\frac{1}{2} + \frac{1}{2}\alpha )\, {\rm J}$
\\
\\
&
$|\Psi_{1,3}\rangle =
 |\sigma\sigma\rangle_1 $
&
E$_{1,3} = (\frac{1}{2} - \frac{1}{2}\alpha )
\,{\rm J}$
\\
\\
&
$|\Psi_{1,2}\rangle =
|(\rho\sigma)_S\rangle $
&
E$_{1,2} = (-\frac{1}{2} + \frac{1}{2}\alpha )
\,{\rm J}$
\\
\\
&
$|\Psi_{1,1}\rangle =
 |(\rho\sigma)_A\rangle $
&
E$_{1,1} = (-\frac{1}{2} - \frac{1}{2}\alpha )
\,{\rm J}$
\\
\\
&
$|\Psi_{0,2}\rangle =
+ \cos(\theta_0) |\sigma\sigma\rangle_0
+ \sin(\theta_0) |\rho\rho\rangle $
&
E$_{0,2} =
(
+ \sqrt{1 - \alpha + \alpha^2}
- \frac{1}{2} - \frac{1}{2}\alpha
)
\,{\rm J}$
\\
\\
&
$|\Psi_{0,1}\rangle =
-\sin(\theta_0) |\sigma\sigma\rangle_0 +
\cos(\theta_0) |\rho\rho\rangle $
&
E$_{0,1} =
(
- \sqrt{1 - \alpha + \alpha^2}
- \frac{1}{2} - \frac{1}{2}\alpha
)
\,{\rm J}$
\\
\\
\label{wavefunctions1}
\end{tabular}
\end{ruledtabular}
\end{table*}

\setcounter{table}{1}

\begin{table*}
\caption{Energy eigenvectors and eigenvalues (cont).}
\vskip 0.5cm
\begin{ruledtabular}
\begin{tabular}{lll}
Spin System
& Eigenvector $|\Psi_{\S_{tot}}\rangle$ (S$_{z\, tot} =$ S$_{\, tot}$)
& Energy
\\
\colrule \hline
&
\\
Linear Tetramer
&
$|\Psi_2\rangle \ \ = \ |\sigma\sigma\rangle_2 $
&
E$_2 = (\frac{1}{2} + \frac{1}{4}\alpha )\, {\rm J}$
\\
\\
&
$|\Psi_{1,3}\rangle =
+ \cos(\theta_1) |\sigma\sigma\rangle_1 +
\sin(\theta_1) |(\rho\sigma)_S\rangle $
&
E$_{1,3} = (+\frac{1}{2}\sqrt{1+\alpha^2} - \frac{1}{4}\alpha )
\,{\rm J}$
\\
\\
&
$|\Psi_{1,2}\rangle =
|(\rho\sigma)_A\rangle $
&
E$_{1,2} = (-\frac{1}{2} + \frac{1}{4}\alpha )
\,{\rm J}$
\\
\\
&
$|\Psi_{1,1}\rangle =
-\sin(\theta_1) |\sigma\sigma\rangle_1 +
\cos(\theta_1) |(\rho\sigma)_S\rangle $
&
E$_{1,1} = (-\frac{1}{2}\sqrt{1+\alpha^2} - \frac{1}{4}\alpha )
\,{\rm J}$
\\
\\
&
$|\Psi_{0,2}\rangle =
+ \cos(\theta_0) |\sigma\sigma\rangle_0
+ \sin(\theta_0) |\rho\rho\rangle $
&
E$_{0,2} =
(
+ \sqrt{1 - \frac{1}{2}\alpha + \frac{1}{4} \alpha^2}
- \frac{1}{2} - \frac{1}{4}\alpha
)
\,{\rm J}$
\\
\\
&
$|\Psi_{0,1}\rangle =
-\sin(\theta_0) |\sigma\sigma\rangle_0 +
\cos(\theta_0) |\rho\rho\rangle $
&
E$_{0,1} =
(
- \sqrt{1 - \frac{1}{2}\alpha + \frac{1}{4} \alpha^2}
- \frac{1}{2} - \frac{1}{4}\alpha
)
\,{\rm J}$
\\
\\
\label{wavefunctions2}
\end{tabular}
\end{ruledtabular}
\end{table*}

\begin{table} [h]
\caption{Inelastic Neutron Scattering Transitions$^a$}
\begin{ruledtabular}
\begin{tabular}{lrlc}

System & & Transition & $\Delta E$ \\

\colrule \hline
& &  \\
Dimer
& I.
&
$|\Psi_0\rangle \to |\Psi_1\rangle$ & $\J$ \\
&&\\
\colrule \hline
&&\\
Symmetric Trimer
& I.
&
$|\Psi_{\frac{1}{2},1}\rangle\to |\Psi_{\frac{3}{2}}\rangle$
& $\frac{3}{2}\,\J$ \\
&
II.
&
$|\Psi_{\frac{1}{2},2}\rangle\to |\Psi_{\frac{3}{2}}\rangle$
&  \\
&&\\
\colrule
&&\\
Isosceles Trimer
&
I.
&
$|\Psi_{\frac{1}{2},1}\rangle \to |\Psi_{\frac{3}{2}}\rangle$
&
$(1+\frac{1}{2}\alpha)\,\J$ \\
&
II.
&
$|\Psi_{\frac{1}{2},2}\rangle \to |\Psi_{\frac{3}{2}}\rangle$
& $\frac{3}{2}\alpha \,\J $ \\
&
III.
&
$|\Psi_{\frac{1}{2},1}\rangle \to |\Psi_{\frac{1}{2},2}\rangle$
&
$(1-\alpha)\,\J$ \\
&&\\
\colrule
&&\\
General Trimer
&
I.
&
$|\Psi_{\frac{1}{2},1}\rangle\to|\Psi_{\frac{3}{2}}\rangle$
&
$(\frac{1}{2}(1+\alpha_s) + \frac{1}{4}f_0 )\,\J$ \\
&
II.
&
$|\Psi_{\frac{1}{2},2}\rangle\to|\Psi_{\frac{3}{2}}\rangle$
&
$(\frac{1}{2}(1+\alpha_s) - \frac{1}{4}f_0 )\,\J$ \\
&
III.
&
$|\Psi_{\frac{1}{2},1}\rangle\to|\Psi_{\frac{1}{2},2}\rangle$
&
$\frac{1}{2}f_0\,\J$ \\
&&\\
\colrule \hline
&&\\
Tetrahedron
& I.
& $|\Psi_{0,1}\rangle \to |\Psi_{1,1}\rangle$ & $\J$ \\
& II.
& $|\Psi_{0,1}\rangle \to |\Psi_{1,2}\rangle$ &  \\
& III.
& $|\Psi_{0,1}\rangle \to |\Psi_{1,3}\rangle$ &  \\
& IV.
& $|\Psi_{0,2}\rangle \to |\Psi_{1,1}\rangle$ &  \\
& V.
& $|\Psi_{0,2}\rangle \to |\Psi_{1,2}\rangle$ &  \\
& VI.
& $|\Psi_{0,2}\rangle \to |\Psi_{1,3}\rangle$ &  \\
& VII.
& $|\Psi_{1,1}\rangle \to |\Psi_{2}\rangle$ & $2\J$ \\
& VIII.
& $|\Psi_{1,2}\rangle \to |\Psi_{2}\rangle$ &  \\
& IX.
& $|\Psi_{1,3}\rangle \to |\Psi_{2}\rangle$ &  \\
&&\\
\colrule
&&\\
Rectangular Tetramer
& I.
& $|\Psi_{0,1}\rangle \to |\Psi_{1,1}\rangle $ & $f_1 \,\J$ \\
& II.
& $|\Psi_{0,1}\rangle \to |\Psi_{1,2}\rangle $ & $(f_1+\alpha)\,\J$ \\
& III.
& $|\Psi_{0,1}\rangle \to |\Psi_{1,3}\rangle $ & $(f_1+1) \,\J$ \\
& IV.
& $|\Psi_{1,1}\rangle \to |\Psi_{0,2}\rangle $ & $f_1 \,\J$ \\
& V.
& $|\Psi_{1,1}\rangle \to |\Psi_{1,2}\rangle $ & $\alpha \,\J$ \\
& VI.
& $|\Psi_{1,1}\rangle \to |\Psi_{1,3}\rangle $ & $\J$ \\
& VII.
& $|\Psi_{1,1}\rangle \to |\Psi_{2}\rangle $ & $(1+\alpha)\,\J$ \\
& VIII.
& $|\Psi_{1,2}\rangle \to |\Psi_{0,2}\rangle $ & $(f_1 - \alpha)\, \J$ \\
& IX.
& $|\Psi_{1,2}\rangle \to |\Psi_{1,3}\rangle $ & $(1-\alpha)\, \J$ \\
& X.
& $|\Psi_{1,2}\rangle \to |\Psi_{2}\rangle $ & $\J$ \\
& XI.
& $|\Psi_{0,2}\rangle \to |\Psi_{1,3}\rangle $ & $ (1-f_1)\,\J$ \\
& XII.
& $|\Psi_{1,3}\rangle \to |\Psi_{2}\rangle $ & $ \alpha \,\J$ \\
&&\\
\colrule &&\\
Linear Tetramer
& I.
& $|\Psi_{0,1}\rangle \to |\Psi_{1,1}\rangle $
& $(f_2 - \frac{1}{2}(f_3-1) ) \,\J$ \\
& II.
& $|\Psi_{0,1}\rangle \to |\Psi_{1,2}\rangle $
& $(f_2 + \frac{1}{2}\alpha ) \,\J$ \\
& III.
& $|\Psi_{0,1}\rangle \to |\Psi_{1,3}\rangle $
& $(f_2 + \frac{1}{2}(f_3+1)) \,\J$ \\
& IV.
& $|\Psi_{1,1}\rangle \to |\Psi_{0,2}\rangle $
& $(f_2 + \frac{1}{2}(f_3-1)) \,\J$ \\
& V.
& $|\Psi_{1,1}\rangle \to |\Psi_{1,2}\rangle $
& $\frac{1}{2}(f_3 - 1 + \alpha )\,\J$ \\
& VI.
& $|\Psi_{1,1}\rangle \to |\Psi_{1,3}\rangle $
& $f_3 \,\J$ \\
& VII.
& $|\Psi_{1,1}\rangle \to |\Psi_{2}\rangle $
& $\frac{1}{2}(f_3 + 1 + \alpha )\,\J$ \\
& VIII.
& $|\Psi_{1,2}\rangle \to |\Psi_{0,2}\rangle $
& $(f_2 - \frac{1}{2}\alpha  )\,\J$ \\
& IX.
& $|\Psi_{1,2}\rangle \to |\Psi_{1,3}\rangle $
& $\frac{1}{2}(f_3 + 1 - \alpha )\,\J$ \\
& X.
& $|\Psi_{1,2}\rangle \to |\Psi_{2}\rangle $
& $\J$ \\
& XI.
& $|\Psi_{0,2}\rangle \to |\Psi_{1,3}\rangle $
& $ (-f_2 + \frac{1}{2}(f_3+1))\,\J$ \\
& XII.
& $|\Psi_{1,3}\rangle \to |\Psi_{2}\rangle $
& $ \frac{1}{2}(-f_3 + 1 + \alpha )\,\J$ \\
\label{INSE}
\end{tabular}
\end{ruledtabular}
\footnotemark[1] {This table uses the abbreviations
$f_0 = \sqrt{(2-\alpha_s)^2 + 3\alpha_d^2}$,
$f_1 = \sqrt{1-\alpha+\alpha^2}$,
$f_2 = \sqrt{1-\alpha/2+\alpha^2/4}$,
$f_3 = \sqrt{1+\alpha^2}$.
}
\end{table}

\begin{table*} [t]
\caption{Specific Heats$^a$}
\begin{ruledtabular}
\begin{tabular}{ll}
& \\
Spin System
&
$C / k_B $
\\
& \\
\colrule \hline
&
\\
Dimer
&
$
3(\bJ)^2 e^{\bJ} /(3+ e^{\bJ})^2
$
\\
& \\ \colrule  \hline & \\
Symmetric Trimer
&
$
\frac{9}{4} (\bJ)^2
e^{\frac{3}{2}\bJ}
/
\big(1 + e^{\frac{3}{2}\bJ}\big)^2
$
\\
& \\ \colrule & \\
Isosceles Trimer
&
$
\frac{1}{2} (\bJ)^2
\Big(
2(1-\alpha)^2 e^{(1+2\alpha)\bJ}
+
(2+\alpha)^2 e^{(1+\frac{1}{2}\alpha)\bJ}
+
9 \alpha^2 e^{\frac{3}{2}\alpha\bJ}
\Big)
/
\Big(
2 +
e^{\frac{3}{2}\alpha\bJ} +
e^{(1+\frac{1}{2}\alpha)\bJ}
\Big)^2
$
\\
& \\ \colrule & \\
General Trimer
&
$
\frac{1}{16} (\bJ)^2 e^{\frac{1}{2}(1+\alpha_s)\bJ}
\Big(
f_0^2 e^{\frac{1}{2}(1+\alpha_s)\bJ}
+
\big( 4 (1+\alpha_s)^2 + f_0^2 \big) \cosh(f_0\bJ/4)
+
4f_0(1+\alpha_s) \sinh(f_0\bJ/4)
\Big) /
$
\\
&
$
\Big(
1 + e^{\frac{1}{2}(1+\alpha_s)\bJ}\cosh(f_0\bJ/4)
\Big)^2
$
\\
& \\ \colrule \hline & \\
Tetrahedron
&
$
18 (\bJ)^2
\Big(10 e^{2\bJ} + 5 e^{3\bJ} + e^{5\bJ} \Big)
/
\Big( 5 + 9 e^{2\bJ} + 2 e^{3\bJ} \Big)^2
$
\\
\label{C_table}
\end{tabular}
\end{ruledtabular}
\footnotemark[1] {This table uses the abbreviation
$f_0 = \sqrt{(2-\alpha_s)^2 + 3\alpha_d^2}$.
}
\end{table*}

\begin{table*} [t]
\caption{Susceptibilities$^a$}
\begin{ruledtabular}
\begin{tabular}{ll}
& \\
Spin System
&
$\chi / (g\mu_B)^2 $
\\
& \\
\colrule \hline
&
\\
Dimer
&
$
2\beta/\big(3+ e^{\bJ}\big)
$
\\
& \\ \colrule  \hline & \\
Symmetric Trimer
&
$\frac{1}{4} \beta
\big(5 + e^{\frac{3}{2}\bJ}\big)
/
\big(1 + e^{\frac{3}{2}\bJ}\big)
$
\\
& \\ \colrule & \\
Isosceles Trimer
&
$
\frac{1}{4} \beta
{ \Big( 10 + e^{\frac{3}{2}\alpha\bJ} + e^{(1+\frac{1}{2}\alpha)\bJ} \Big) }
/
{\Big(2 + e^{\frac{3}{2}\alpha \bJ} + e^{(1+\frac{1}{2}\alpha)\bJ} \Big) }
$
\\
& \\ \colrule & \\
General Trimer
&
$ \frac{1}{4} \beta
{ \Big( 5 + e^{\frac{1}{2}(1+\alpha_s) \bJ}\cosh( f_0\bJ/4) \Big) }
/
{ \Big( 1 + e^{\frac{1}{2}(1+\alpha_s) \bJ}\cosh(f_0\bJ/4) \Big) }
$
\\
& \\ \colrule \hline & \\
Tetrahedron
&
$
2\beta \Big(5 + 3 e^{2\bJ}\Big)
/
\Big(5 + 9 e^{2\bJ} + 2 e^{3\bJ}\Big)
$
\\
& \\ \colrule & \\
Rectangular Tetramer
&
$
2\beta \Big(5 + e^{\bJ} + e^{\alpha\bJ} + e^{(1+\alpha)\bJ}\Big)
/
\Big(
5 + 3 e^{\bJ} + 3 e^{\alpha\bJ} + 3 e^{(1+\alpha)\bJ}
+ 2 e^{(1+\alpha)\bJ}\cosh(f_1\bJ/2)
\Big)
$
\\
& \\ \colrule  & \\
Linear Tetramer
&
$
2\beta
\Big(5 + e^{\bJ} +
2 e^{\frac{1}{2}(1+\alpha)\bJ}\cosh(f_3\bJ/2)
\Big)
/
$
\\
&
$
\Big(
5 + 3 e^{\bJ} +
2 e^{(1+\frac{1}{2} \alpha)\bJ}\cosh(f_2\bJ)
+
6 e^{\frac{1}{2}(1+\alpha)\bJ}\cosh(f_3\bJ/2)
\Big)
$
\\
\label{chi_table}
\end{tabular}
\end{ruledtabular}
\footnotemark[1] {This table uses the abbreviations
$f_0 = \sqrt{(2-\alpha_s)^2 + 3\alpha_d^2}$,
$f_1 = \sqrt{1-\alpha+\alpha^2}$,
$f_2 = \sqrt{1-\alpha/2+\alpha^2/4}$,
$f_3 = \sqrt{1+\alpha^2}$.
}
\end{table*}

\end{document}